\documentclass{article}

\PassOptionsToPackage{numbers, compress}{natbib}
\usepackage[preprint]{neurips_2026}

\usepackage[utf8]{inputenc} % allow utf-8 input
\usepackage[T1]{fontenc}    % use 8-bit T1 fonts
\usepackage{hyperref}       % hyperlinks
\usepackage{url}            % simple URL typesetting
\usepackage{booktabs}       % professional-quality tables
\usepackage{amsfonts}       % blackboard math symbols
\usepackage{nicefrac}       % compact symbols for 1/2, etc.
\usepackage{microtype}      % microtypography
\usepackage[table,xcdraw]{xcolor} % colors

\usepackage{amsmath}
\usepackage{amssymb}
\usepackage{mathtools}
\usepackage{amsthm}
\usepackage{listings}
\usepackage[most]{tcolorbox} % 引入tcolorbox用于美化方框
\usepackage{enumitem}        % 用于紧凑列表
\tcbuselibrary{breakable}    % 允许方框跨页
\usepackage{graphicx}
\usepackage{subcaption}
\usepackage{multirow}

\usepackage[capitalize,noabbrev]{cleveref}

\theoremstyle{plain}

\theoremstyle{definition}

\theoremstyle{remark}

\usepackage[textsize=tiny]{todonotes}

\title{Q-Probe: Scaling Image Quality Assessment to High Resolution via Context-Aware Agentic Probing}
\vspace{-8mm}
\author{%
  \vspace{0mm}
  Xiang Li$^{1,2,*}$ \quad
  Xueheng Li$^{1,2,*}$ \quad
  Yu Wang$^{2,1*}$ \quad
  Xuanhua He$^{3}$ \\
  Zhangchi Hu$^{1}$ \quad
  Chengjun Xie$^{2}$ \\
  \vspace{-4mm} \\
  $^1$University of Science and Technology of China \quad \\
  $^2$Hefei Institutes of Physical Science, Chinese Academy of Sciences \\
  $^3$The Hong Kong University of Science and Technology \\
  \vspace{-2mm}
  $^*$These authors contributed equally to this work.
  \vspace{-4mm}
}

\begin{document}
\vspace{-5mm}
\maketitle

\begin{figure}[h]
  \centering
  \includegraphics[width=1.0\textwidth]{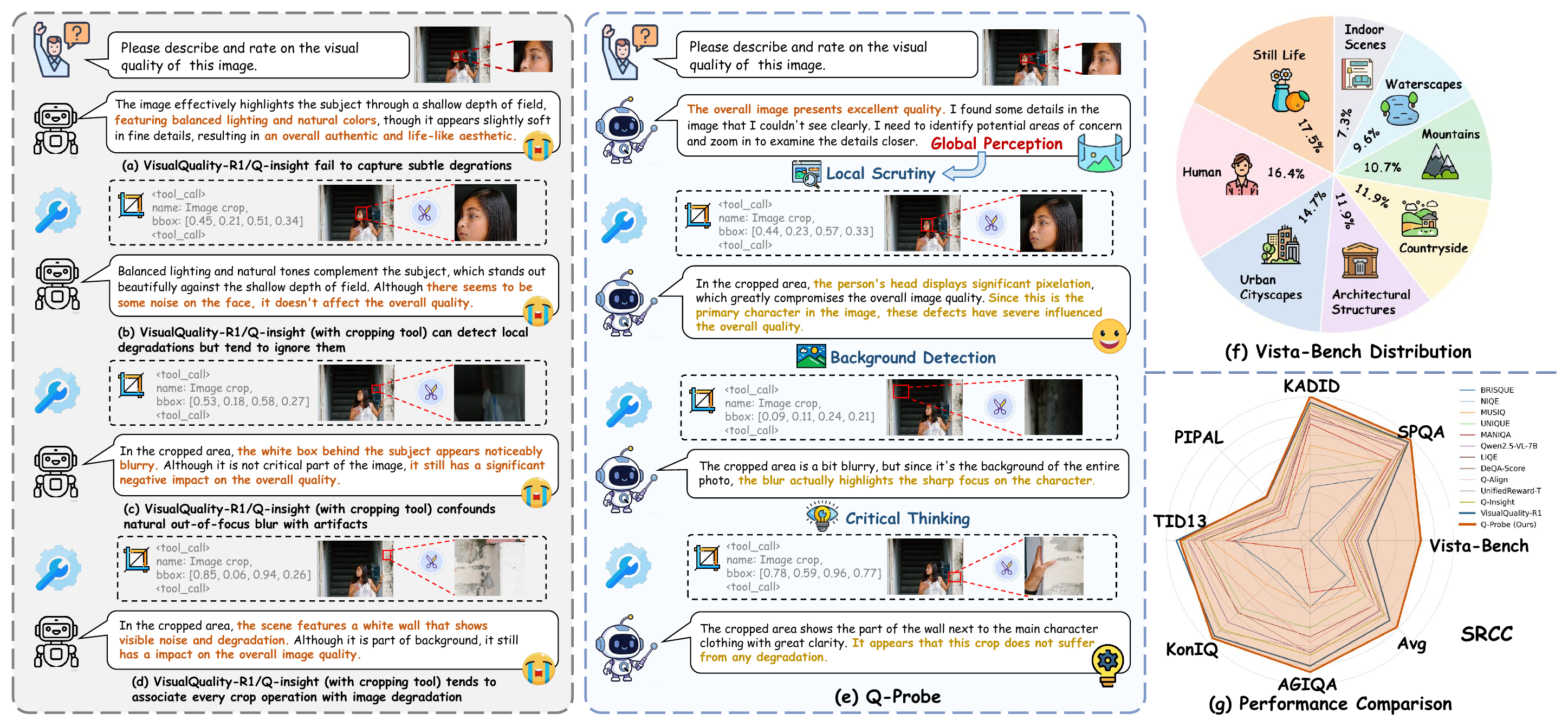} 
  \caption{Challenges in detecting subtle distortions via global perception versus local zooming. (a) Existing MLLMs fail to capture subtle local artifacts. (b) Even when visible via cropping, Semantic Robustness Bias causes models to ignore defects in key semantic areas (e.g., face). (c-d) Naive zooming leads to \textit{Logic Collapse}, where the model misinterprets natural bokeh as blur (c) or falsely learns that Zooming implies Low Quality (d). (e) Q-Probe mimics human active viewing (Global Perception $\rightarrow$ Local Scrutiny $\rightarrow$ Critical Thinking) to correctly distinguish artifacts from natural effects. (f-g) Data distribution of the high-resolution Vista-Bench  and performance comparison showing Q-Probe's superiority.}
  \label{fig:teaser}
  \vspace{-4mm}
\end{figure}
\begin{abstract}
Reinforcement Learning (RL) has empowered Multimodal Large Language Models (MLLMs) to achieve superior human preference alignment in Image Quality Assessment (IQA). However, existing RL-based IQA models typically rely on coarse-grained global views, failing to capture subtle local degradations in high-resolution scenarios. While emerging ``Thinking with Images'' paradigms enable multi-scale visual perception via zoom-in mechanisms, their direct adaptation to IQA induces spurious ``cropping-implies-degradation'' biases and misinterprets natural depth-of-field as artifacts. To address these challenges, we propose Q-Probe, the first agentic IQA framework designed to scale IQA to high resolution via context-aware probing. First, we construct Vista-Bench, a pioneering benchmark tailored for fine-grained local degradation analysis in high-resolution IQA settings. Furthermore, we propose a three-stage training paradigm that progressively aligns the model with human preferences, while simultaneously eliminating causal bias through a novel context-aware cropping strategy. Extensive experiments demonstrate that Q-Probe achieves state-of-the-art performance in high-resolution settings while maintaining superior efficacy across resolution scales.
\end{abstract}

% 将 ICML 的头图 (Teaser Figure) 移到 Abstract 之后

\section{Introduction}

Image Quality Assessment (IQA) serves as a pivotal fundamental technique in computer vision, designed to emulate human perception of visual fidelity \cite{fang2020perceptual, saha2023re, jia2025refine}. Within this domain, No-Reference (NR) methods \cite{wang2021survey, mao2025no} are highly valued in practical scenarios due to their independence from pristine reference images. However, the transition from traditional handcrafted features \cite{ahmed2019image} to deep learning paradigms \cite{talebi2018nima, yu2024sf} has long been plagued by challenges regarding overfitting and out-of-distribution generalization \cite{zhong2024causal}. The emergence of Multimodal Large Language Models (MLLMs) \cite{wang2024qwen2, wang2025internvl3} has provided a novel solution to this dilemma. Through the Chain-of-Thought (CoT) mechanism, MLLMs integrate low-level distortions (e.g., noise, blur) and high-level semantics (e.g., aesthetics, content), markedly improving model generalization \cite{wu2024q, you2025teaching}. However, current Supervised Fine-Tuning (SFT) paradigms are still hindered by high annotation expenses and shallow reasoning, failing to fully unleash the inherent cognitive potential of MLLMs.

Recent research paradigms are shifting towards leveraging Reinforcement Learning (RL) to further align models with human perception \cite{cai2025q, zhao2025reasoning}. This direction advocates for the explicit design of reward functions to unify the complementary dimensions of ``response consistency'' and ``preference alignment,'' thereby overcoming the limitations of traditional approaches that optimize solely for ranking or regression accuracy. Two dominant RL paradigms have emerged in IQA. Regression-based methods (e.g., Q-Insight \cite{li2025q}) treat IQA as absolute score prediction, employing verifiable tolerance rewards to align accuracy with human preference. Conversely, comparison-based methods (e.g., VisualQuality-R1 \cite{wu2025visualquality}) derive quality from relative differences. By optimizing ranking distributions against human preferences, they induce emergent CoT capabilities that balance ranking accuracy with logical interpretability.

Nevertheless, RL-based IQA methods suffer from a heavy reliance on global macroscopic views, making it difficult to precisely capture subtle local degradation features (Fig. \ref{fig:teaser}(a)). Even when such artifacts are detected, they are frequently overlooked due to insufficient attentional allocation. In real-world scenarios, degradation in high-value semantic regions (e.g., faces or license plates) should trigger a severe penalty in the quality metric, whereas existing models frequently overestimate quality in such cases due to semantic robustness bias (Fig. \ref{fig:teaser}(b)). Moreover, these methods are typically restricted to processing global quality. In high-resolution contexts, this coarse-grained approach fails to resolve small objects and subtle local artifacts, leading to marked biases in quality prediction.

The recent emergence of the ``Thinking with Images'' \cite{OpenAI2025Thinking} paradigm has spurred the use of multi-scale tools for high-resolution tiny object perception \cite{zheng2025deepeyes, wang2025pixel}. While this offers a promising avenue for fine-grained perception in IQA, directly transposing this local zoom-in strategy presents severe challenges. On one hand, traditional visual zoom-in methods relying on region-level supervision benefit from the diversity of target objects, merely requiring SFT trajectories to encompass the target regions. In the IQA context, however, training exclusively on degraded regions is prone to inducing overfitting, causing the model to establish a spurious causal correlation that ``cropping implies low quality'' (Fig. \ref{fig:teaser}(d)). Consequently, the model loses its discriminative capability regarding content, exhibiting a systematic low-score bias across all local views. On the other hand, even when equipped with local zoom-in tools to facilitate fine-grained semantic analysis, existing MLLMs often overlook the critical dependency of quality assessment on the global context. Taking the Depth-of-Field (DoF) effect in photography as an example, a sharp foreground contrasted with a blurred background typically signifies high-quality imaging. Yet, when isolated from the holistic view and focused solely on the background, models tend to misinterpret natural optical bokeh as degradation (e.g., blur), thereby erroneously assigning a low quality score (Fig. \ref{fig:teaser}(c)). Furthermore, the current IQA landscape lacks a benchmark dedicated to local fine-grained degradations in high-resolution scenarios, which significantly constrains the evolution of algorithms toward more refined perception.

To address the aforementioned challenges, we propose \textbf{Q-Probe} (Fig. \ref{fig:teaser}(e)), the first agentic IQA model grounded in the Thinking with Images paradigm and designed specifically for high-resolution scenarios. To enable such fine-grained perception given the lack of dedicated benchmarks, we first construct \textbf{Vista-Bench} (Fig. \ref{fig:teaser}(f)), covering a wide range of scenes and local distortion patterns. To ensure the authenticity and precision of the synthesized artifacts, we employ the wavelet transform \cite{zhang2019wavelet} to decouple image structure from texture, selectively injecting degradations into texture-rich regions to simulate realistic impairments. To guarantee authoritative ground truth, we deploy a dedicated panel of six human specialists to perform a hierarchical evaluation, calculating weighted assessments based on local degradation severity and the semantic significance of the regions.

Building on this foundation, we introduce a three-stage training paradigm mimicking the human visual mechanism of ``global perception, local scrutiny,'' enabling Q-Probe to dynamically optimize tool invocation for comprehensive assessment across resolution scales. Initially, we conduct RL pre-training using a subset of low-resolution images to establish a foundational perception of image quality. This phase is crucial for decoupling the overall learning complexity. By first anchoring the model in the fundamental concept of Mean Opinion Scores (MOS), we establish a stable baseline that allows subsequent stages to focus exclusively on mastering intricate agentic behaviors without the burden of basic quality alignment.

Following this, we curate \textbf{Probe-CoT-3K}, a dataset comprising mixed-resolution imagery and reasoning trajectories, to facilitate an SFT warm-up phase. Here, we devise a generation strategy that integrates a global overview with diverse local cropping mechanisms (e.g., degradation capture, clarity localization, and distant view assessment). Specifically, the trajectories simulate an initial global reasoning phase, utilizing a holistic view to determine the necessity of further fine-grained perception. During the local cropping phase, unlike traditional VQA methods that focus solely on target semantics, we mandate that cropped regions encompass all degraded patches while simultaneously preserving areas exhibiting pristine natural DoF or clear foregrounds. This design not only effectively eliminates the spurious causal correlation associating ``local cropping'' with ``low quality'' but also empowers the model to discern the optimal timing of tool invocation, grounded in a balanced integration of global and local perspectives. However, the strategy of incorporating background context to ensure logical robustness inevitably reduces the model's precision in localizing subtle degradations, requiring further optimization. To maximize the model's capacity to identify such fine-grained artifacts, we construct the \textbf{Probe-RL-4K} dataset and implement Reinforcement Fine-Tuning, synchronously optimizing reasoning trajectories and tool invocation precision for accurate decision-making in complex scenarios. Extensive experimental results demonstrate that Q-Probe significantly outperforms existing models in high-resolution settings, accompanied by varying degrees of performance gains in low-resolution scenarios.

Our contributions are summarized as follows:
\begin{itemize}
    \item We propose Q-Probe, the first agentic IQA model for high-resolution scenarios, leveraging adaptive cropping to achieve superior capability in assessing both global quality and subtle local degradations.
    \item We introduce Vista-Bench, the pioneering benchmark dedicated to high-resolution fine-grained degradation analysis, encompassing a broad spectrum of scenes and local distortion patterns.
    \item We construct the Probe-CoT-3K and Probe-RL-4K datasets, leveraging a context-aware cropping strategy to eliminate spurious correlations and generate high-quality CoT trajectories for mastering tool usage.
    \item We devise a progressive three-stage training curriculum that first aligns global perception with human preference, then stabilizes logical reasoning via SFT, and finally leverages decoupled reward guidance to recover localization precision.
\end{itemize}

\section{Related Work}
\label{sec:related_work}

\subsection{Image Quality Assessment}
Image Quality Assessment (IQA) has evolved from early reliance on handcrafted features based on Natural Scene Statistics (NSS)~\cite{mittal2012no, mittal2012making} to data-driven deep learning models. While subsequent CNN and Transformer-based approaches achieved performance gains, they often treated IQA as a black-box regression task. The recent advent of Multimodal Large Language Models (MLLMs) introduced a semantic turn, with methods like {Q-Align}~\cite{wu2023q} and {DeQA-Score}~\cite{you2025teaching} leveraging foundation models for quality scoring via supervised fine-tuning (SFT). However, these SFT-based paradigms rely heavily on expensive annotations and often suffer from shallow reasoning patterns.
To address these limitations, Reinforcement Learning (RL) has emerged as a pivotal mechanism for aligning MLLMs with human preferences. Notably, Q-Insight~\cite{li2025q} pioneered the application of Group Relative Policy Optimization (GRPO) to visual quality understanding, while VisualQuality-R1~\cite{wu2025visualquality} reformulated IQA as a ranking task via RL-to-Rank.

\subsection{Thinking with Images}
The ``Thinking with Images'' paradigm, introduced by OpenAI-o3~\cite{OpenAI2025Thinking}, has spurred the development of agentic visual reasoning MLLMs that interleave image and text reasoning with iterative visual analysis. {DeepEyes}~\cite{zheng2025deepeyes} provides an open-source implementation, demonstrating that end-to-end RL can incentivize models to adopt this behavior for fine-grained tasks. However, subsequent works reveal that pure RL is insufficient for complex, multi-turn interactions. {Pixel Reasoner}~\cite{wang2025pixel} identifies a critical ``learning trap'' where models bypass nascent visual tools, proposing a two-phase approach: a cold-start phase to establish tool use followed by curiosity-driven RL. Similarly, other works have adopted multi-stage training to activate deep trajectories for hard visual searches.
Despite these advances, directly transplanting this agentic probing to IQA remains non-trivial. Unlike object detection, cropping in IQA introduces a {causal ambiguity}: models trained on cropped degradation patches tend to overfit, learning a spurious correlation that ``zooming implies degradation,'' thereby penalizing high-quality crops. 
{Q-Probe} addresses this gap by introducing a context-aware probing mechanism. 
\begin{figure*}[t]
  \begin{center}
    % Placeholder for HRIQA-Bench Pipeline Image
    \includegraphics[width=\textwidth]{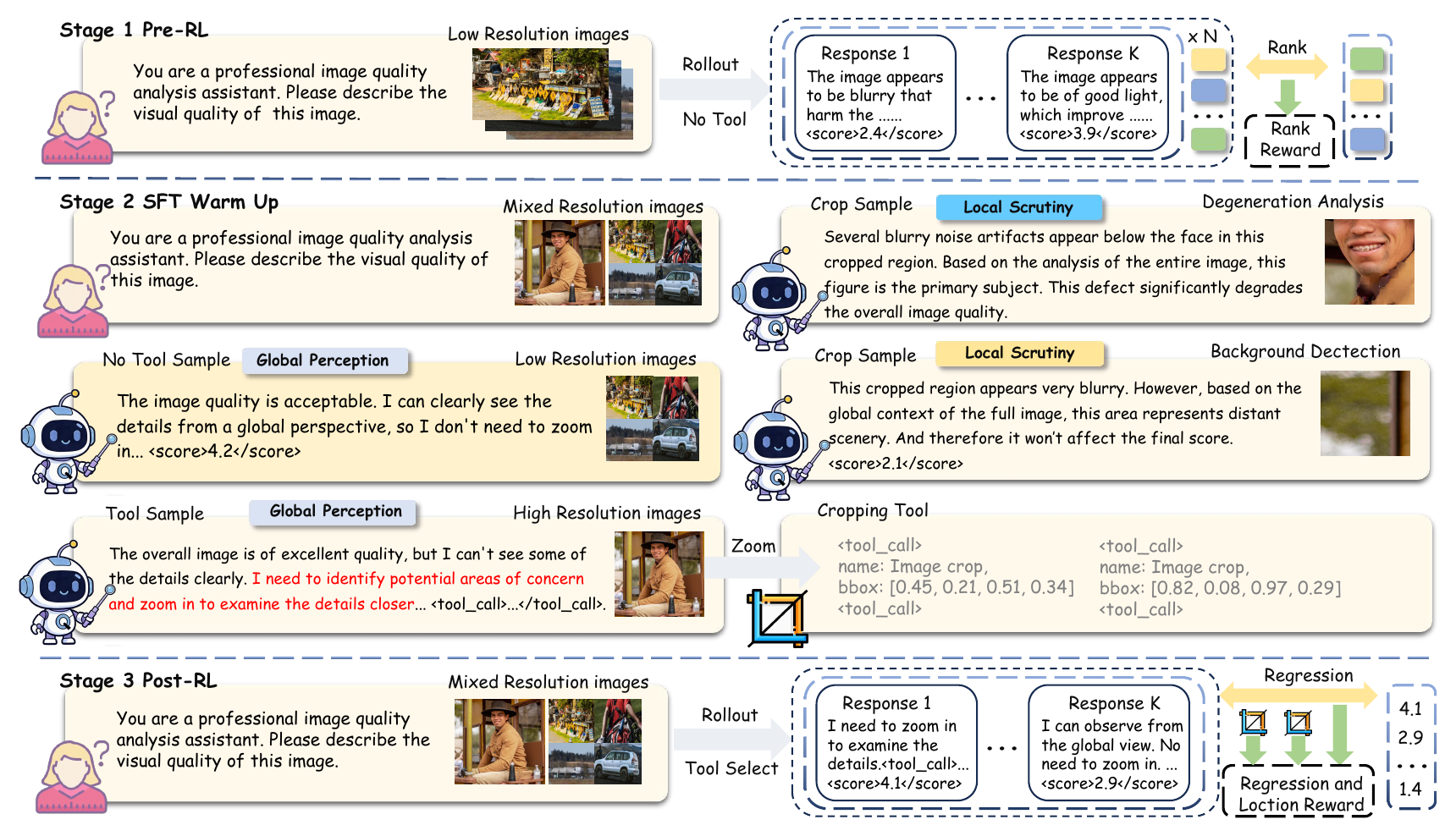} 
    \caption{Overview of the three-stage training framework. Initially, RL Pre-training leverages ranking rewards to align global perception with human preferences. Subsequently, hybrid-resolution SFT enables the model to acquire robust logical reasoning. Finally, the RL Post-training stage fine-tunes the model for precise degradation detection and adaptive tool invocation.}
    \label{fig:benchmark}
  \end{center}
  \vspace{-4mm}
\end{figure*}
\begin{figure*}[t]
  \centering
  % Placeholder for Figure 3: Data Flywheel
  \includegraphics[width=\textwidth]{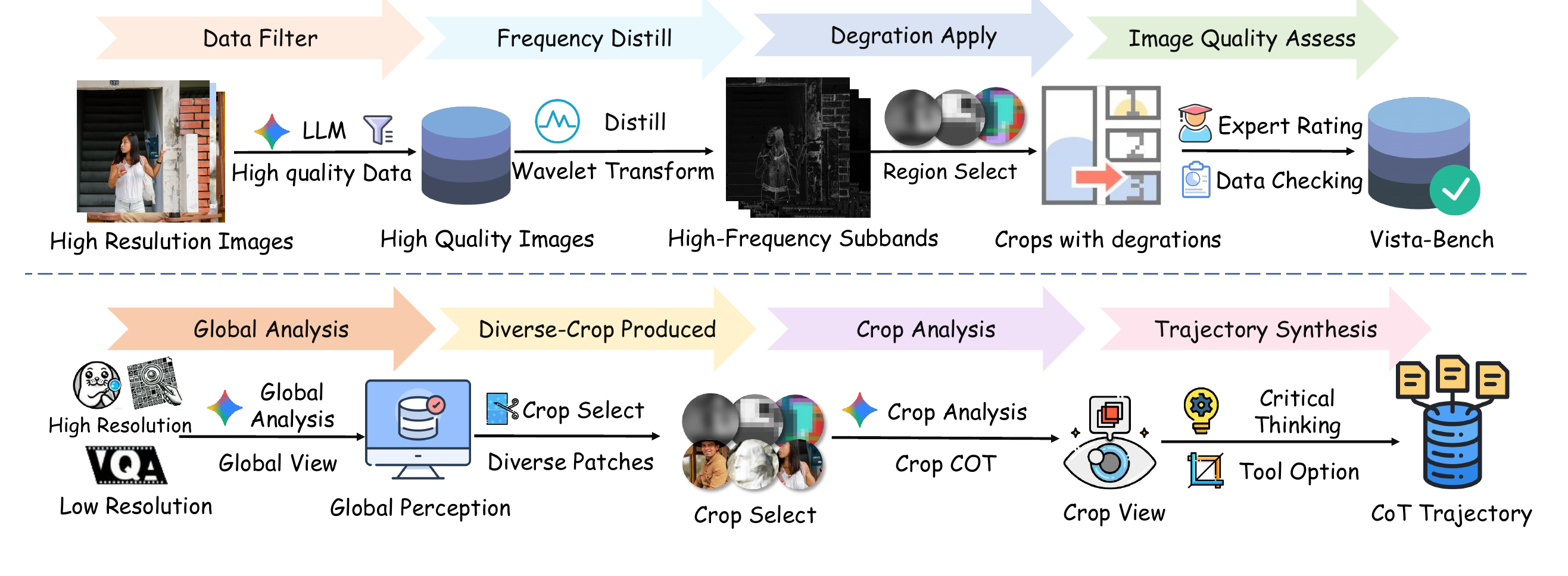} 
  \caption{This diagram illustrates the construction pipeline of Vista-Bench and the Data Flywheel for SFT. Specifically, we utilize wavelet transforms to decouple structure from texture, selectively injecting artifacts into texture-rich semantic regions. To ensure annotation accuracy, we employ a panel of domain experts to manually generate importance-weighted scores for fine-grained perception probing. To support SFT, we generate traces that interleave global overviews, defect zooming, and context verification (scrutinizing clear regions), thereby preventing the model from associating tool usage solely with defects.}
  \label{fig:datapipeline}
  \vspace{-4mm}
\end{figure*}

\section{Methodology}
\label{sec:method}

\subsection{Overview}
\label{sec:overview}

Q-Probe emulates the human ``coarse-to-fine'' visual mechanism via a progressive three-stage curriculum, balancing global aesthetic perception with local defect scrutiny. To facilitate this, we constructed {Vista-Bench}, a high-resolution benchmark derived from public datasets through wavelet-based artifact injection, manual scoring by an expert panel. The training pipeline of Q-probe begins with perception alignment using Group Relative Policy Optimization (GRPO) \cite{guo2025deepseek} to align with the human concept of Mean Opinion Scores (MOS). It then progresses to a hybrid-resolution SFT phase driven by a Data Flywheel, employing context-aware cropping to dissociate tool usage from negative quality bias. Finally, a decoupled RL post-training stage utilizes a bifurcated reward mechanism to jointly optimize defect localization and scoring, resolving the conflict between exploration diversity and detection precision. Details are shown in Fig. \ref{fig:benchmark}.

\subsection{Data Construction: Vista-Bench}
\label{sec:data_construction}

Existing datasets usually consist of low-resolution images ($<1$K) where degradations are typically applied globally rather than to localized regions. To evaluate fine-grained perception in the high-resolution quality assessment era, we constructed the \textbf{Vista-Bench}.

The dataset is curated from high-quality samples within the {HR-Bench 4K~\cite{wang2025divide}}, {Pixel-Reasoner~\cite{wang2025pixel}} and {UHD-IQA}~\cite{hosu2024uhd} datasets. As shown in Fig. \ref{fig:datapipeline}, the construction pipeline proceeds as follows: we first utilize wavelet transforms to extract global high-frequency components and apply specific degradations (e.g., blur,  compression, mosaic). This approach ensures that the degradation is applied to the semantic foreground regions as much as possible. Subsequently, we employ a dedicated panel of six human experts to conduct a rigorous, two-stage manual annotation. The evaluators first assess the global visual quality, and then scrutinize the explicitly cropped regions to generate fine-grained quality scores based on the severity of the local degradation and the semantic importance of the region. The resulting benchmark encompasses over 3,000 high-resolution ($>4$K) images, characterized by authoritative, multi-domain, and locally annotated degradations. We provide more details in Appendix \cref{sec:appendix_data_construction}.

\subsection{Stage 1: Perception Alignment via Pre-RL}
\label{sec:stage1}

\textbf{Why Pre-alignment is Required.} Before introducing complex tool usage, we strategically decouple the learning objectives to prevent the model from being overwhelmed by simultaneous tasks. Requiring the policy to concurrently learn whether to zoom in, where to zoom in, how to weigh the importance of local crops against the global image, and the fundamental concept of Mean Opinion Scores (MOS) makes the optimization highly unstable. Therefore, we employ a dataset of 3,000 low-resolution images randomly sampled from the KADID-10k \cite{lin2019kadid} dataset to train the model via a pairwise variant of GRPO. This initial stage explicitly isolates and anchors the learning of the MOS concept and foundational aesthetic perception. Consequently, Stage 2 can focus exclusively on mastering the three intricate crop-related capabilities without the confounding burden of basic quality alignment.

\textbf{Why not use SFT as Stage 1.} A fundamental innovation of our Q-Probe framework lies in its pioneering three-stage training curriculum: GRPO $\rightarrow$ SFT $\rightarrow$ GRPO. To the best of our knowledge, this is the first framework to adopt such an interleaved reinforcement learning paradigm for multimodal visual quality assessment. We observed that employing SFT as Stage 1 significantly entrenches the model's fixation on global quality analysis, rendering it exceedingly difficult to learn the relevant zoom-in invocation capabilities during the second stage.
Crucially, utilizing standard SFT to directly instill the concept of Mean Opinion Scores (MOS) inevitably forces the model to anchor its judgments on macroscopic, global-view heuristics. This deepens the ``Semantic Robustness Bias,'' causing the model's cognitive process to be disproportionately dominated by holistic image semantics rather than objective local fidelity. Once trapped in this global bias, the model exhibits severe resistance during Stage 2; it struggles to learn the intricate mechanics of executing cropping actions, analyzing subtle localized defects, and reconciling these micro-imperfections with the overall global context. For more extended discussion, please refer to Appendix \cref{subsec:pairwise}.

% By instead deploying GRPO driven by pairwise relative quality rewards in the initial stage, we circumvent this limitation. The relative ranking reward endows the model with a robust, foundational understanding of the MOS concept without overfitting to a rigid, global-only perspective. This strategy effectively limits the over-reliance on global semantics, preserving the model's exploratory plasticity. Consequently, it establishes an optimal, unbiased foundation for the subsequent hybrid-resolution SFT stage to truly master fine-grained, context-aware agentic probing.

\textbf{Probabilistic Ranking via Thurstone Model.} Unlike standard regression approaches that treat quality scores as deterministic point estimates, we model visual quality as an intrinsically relative concept governed by the Thurstone model \cite{thurstone2017law} case V. For a training pair of images $\{x_i, x_j\}$ (effectively a batch size of $B=2$), the policy model $\pi_\theta$ generates $K$ distinct reasoning paths and quality scores for each image: $q(x_i) = [q_1(x_i), \dots, q_K(x_i)]^\intercal$. This distribution inherently captures predictive uncertainty,  which serves as a cornerstone for reliable ranking.

We calculate the asymmetric comparative probability that $x_i$ is perceptually superior to $x_j$ by standardizing the difference in their predicted means against their joint uncertainty. To improve numerical stability, we first define the standardized difference score $\mathcal{Z}_{ij}^k$:
\begin{equation}
    \mathcal{Z}_{ij}^k = \frac{q_k(x_i) - \mu(q(x_j))}{\sqrt{\sigma^2(q(x_i)) + \sigma^2(q(x_j)) + \gamma}}
\end{equation}

where $\mu(\cdot)$ and $\sigma^2(\cdot)$ denote the sample mean and variance of the predicted scores group, $q_k(x_i)$ represents the $k$-th score estimate for image $x_i$, and $\gamma$ is a small constant ensuring numerical stability. The comparative probability is then derived via the standard Gaussian Cumulative Distribution Function (CDF):
\begin{equation}
    P_k(x_i \succ x_j) = \Phi\left( \mathcal{Z}_{ij}^k \right)
\end{equation}

where $\Phi(\cdot)$ is the Gaussian CDF, representing the probability $P_k$ that image $x_i$ is perceptually superior to $x_j$. Crucially, we explicitly leverage the sample variances derived from the GRPO group outputs to dynamically accommodate predictive uncertainty, preventing over-confident errors on ambiguous images.

\textbf{Fidelity Reward and Optimization.}
To rapidly align the model's predictions with human perception, we define the ground truth preference $y_{ij}$ based on Mean Opinion Scores (MOS): $y_{ij} = 1$ if $\text{MOS}(x_i) > \text{MOS}(x_j)$, $0.5$ if equal, and $0$ otherwise. The optimization is driven by a continuous fidelity reward. We define the ranking reward $R_{rank}$ as:

\begin{small}
\begin{equation}
\begin{split}
    R_{rank}(x_i) =\ & y_{ij} \cdot P_k(x_i \succ x_j) \\
    & + (1 - y_{ij}) \cdot (1 - P_k(x_i \succ x_j))
\end{split}
\end{equation}
\end{small}

where $y_{ij} \in \{0, 0.5, 1\}$ represents the ground truth preference label derived from human ratings. This reward is designed to capture fine-grained distinctions in quality ranking. To ensure training stability, we maximize the GRPO objective $\mathcal{J}(\theta)$, which integrates the clipped surrogate advantage and a KL-divergence penalty:

\begin{small}
\begin{align}
\label{eq:grpo_loss}
    \mathcal{J}(\theta) &= \mathbb{E}\Bigg[ \min\left( \frac{\pi_\theta}{\pi_{old}} A_k, \text{clip}\left(\frac{\pi_\theta}{\pi_{old}}, 1-\epsilon, 1+\epsilon\right) A_k \right) \nonumber \\
    &\quad - \beta D_{KL}(\pi_\theta || \pi_{ref}) \Bigg]
\end{align}
\end{small}
where $\pi_\theta$ and $\pi_{old}$ represent the current and previous policies respectively, $A_k$ is the advantage score computed from reward signals, $\epsilon$ is the clipping threshold to constrain policy updates, $\pi_{ref}$ is the reference policy (typically the initial SFT model), and $\beta$ is a coefficient controlling the strength of the KL-divergence penalty $D_{KL}$. This pre-alignment stage establishes a solid baseline for global image understanding, converging towards the mechanisms of human MOS.

\subsection{Stage 2: Hybrid-Resolution SFT via Data Flywheel}
\label{sec:stage2}

To bridge the gap between global perception and local scrutiny, we construct the \textbf{Probe-CoT-3K} dataset using a Data Flywheel approach (as illustrated in Fig. \ref{fig:datapipeline}). The dataset follows a 2:1 ratio of low-to-high resolution images.

\textbf{The SFT Dilemma: Precision vs. Diversity.} 
In traditional Visual Question Answering (VQA), crop content inherently exhibits high diversity. However, in Agentic IQA, if SFT data is invariably centered on degradations, the model overfits to a biased mode: \textit{``Tool Usage implies Degradation and Low Quality,''} resulting in a fundamental breakdown of logical reasoning. 

\textbf{Context-Aware Trajectory Generation.}
To mitigate this, we synthesize CoT trajectories encompassing genuine degradations, natural depth-of-field, and pristine foregrounds. This strategy effectively severs the spurious causal correlation associating ``cropping'' with ``low quality''. 

Specifically, the SFT data primarily includes two types. First, the model can directly observe the global details and does not need to zoom in. Second, for high-resolution images or when global details are unclear, the model decides to zoom in for further analysis. After zooming in, if a defect is discovered, the model applies a penalty based on the location's importance to the overall image; however, if no defects are found, or if the model reasons that the observed blurriness is simply a normal out-of-focus background (depth of field), it decides not to penalize.

We optimize the model using the standard cross-entropy loss over the mixed sequence of reasoning and action tokens $y$:
\begin{equation}
    \mathcal{L}_{SFT} = - \sum_{t=1}^{L} \log P(y_t | y_{<t}, \mathbf{x})
\end{equation}
where $\mathcal{L}_{SFT}$ is the SFT loss, $L$ denotes the sequence length, $\mathbf{x}$ represents the multimodal input (image and instruction), and $y_t$ is the $t$-th token in the generated target sequence $y$ given the preceding tokens $y_{<t}$. Following the SFT phase, the model establishes a robust understanding of the global image context and acquires the capability to accurately discern diverse local regional features. However, intentionally introducing a zero-penalty region negatively impacts the model's zoom-in accuracy. Experimental results regarding this can be found in Appendix \cref{tradeoff}.

\subsection{Stage 3: Precision Pursuit via Decoupled Post-RL}
\label{sec:stage3}

As the SFT phase strategically trades localization precision for logical robustness, we construct and leverage the \textbf{Probe-RL-4K} dataset in the final stage to further refine the model's fine-grained detection capabilities.

\textbf{Decoupled Reward Mechanism.}
We implement a decoupled reward mechanism that disentangles the optimization of the ``Looking Policy'' from the ``Scoring Policy.'' We independently define the accuracy reward $R_{acc}$ and the localization reward $R_{loc}$:
\begin{align}
    R_{acc} &= \exp\left( - \frac{|s_{pred} - s_{MOS}|}{\tau} \right) \\
    R_{loc} &= \mathbb{I}(has\_defect) \cdot \text{IoU}(B_{pred}, B_{gt})
\end{align}
where $s_{pred}$ and $s_{MOS}$ denote the predicted score and ground truth MOS respectively, and $\tau$ is a temperature hyperparameter controlling the sensitivity of the accuracy reward. For localization, $\mathbb{I}(\cdot)$ is an indicator function that equals 1 if a defect exists in the image, and $\text{IoU}(\cdot)$ computes the Intersection over Union between the predicted crop box $B_{pred}$ and the ground truth defect region $B_{gt}$. This reward incentivizes the model to accurately capture degradations and precisely deploy tools for regional magnification.

The total reward $R_{total}$ is then composed as:
\begin{equation}
    R_{total} = \underbrace{\alpha R_{acc}}_{\text{Scoring}} + \underbrace{\beta R_{loc}}_{\text{Looking}} + \gamma R_{format}
\end{equation}

where $\alpha$, $\beta$, and $\gamma$ are hyperparameters weighting the contribution of the accuracy reward, localization reward, and format compliance reward $R_{format}$, respectively. This decoupled reward optimizes the model to refine its localization capabilities ($\beta R_{loc}$), building upon the robust logic reasoning capability acquired during SFT. 

\begin{table*}[t!]
\centering
\renewcommand{\arraystretch}{1.1} 
\setlength{\tabcolsep}{3pt}

\caption{Comprehensive performance comparison (SRCC / PLCC) on standard and high-resolution datasets. \textbf{Q-Probe} achieves SOTA results, especially on the high-resolution \textbf{Vista} benchmark. Methods are color-coded by category: \colorbox{gray!10}{Handcrafted}, \colorbox{blue!5}{Deep Learning}, and \colorbox{green!5}{MLLMs-based}.}
\label{tab:main_results_combined}

\resizebox{0.95\textwidth}{!}{%
    \begin{tabular}{l|ccccccc|c}
    \toprule
    \textbf{Methods} & \textbf{Vista} & \textbf{SPAQ} & \textbf{KADID} & \textbf{PIPAL} & \textbf{TID13} & \textbf{KonIQ} & \textbf{AGIQA} & \textbf{Avg} \\
    \midrule
    \multicolumn{9}{c}{\textbf{Spearman Rank Correlation Coefficient (SRCC)}} \\ 
    \midrule
    \rowcolor{gray!10}
    BRISQUE~\cite{mittal2012no} & 0.211 & 0.614 & 0.429 & 0.242 & 0.548 & 0.385 & 0.497 & 0.418 \\
    \rowcolor{gray!10}
    NIQE~\cite{mittal2012making}    & 0.247 & 0.676 & 0.487 & 0.357 & 0.532 & 0.421 & 0.533 & 0.465 \\
    \midrule
    \rowcolor{blue!5}
    MUSIQ~\cite{ke2021musiq}   & 0.357 & 0.720 & 0.647 & 0.317 & 0.670 & 0.473 & 0.494 & 0.525 \\
    \rowcolor{blue!5}
    UNIQUE~\cite{zhang2021uncertainty}  & 0.373 & 0.751 & 0.513 & 0.393 & 0.703 & 0.649 & 0.608 & 0.570 \\
    \rowcolor{blue!5}
    MANIQA~\cite{yang2022maniqa}  & 0.388 & 0.745 & 0.760 & 0.338 & 0.589 & 0.213 & 0.422 & 0.494 \\
    \midrule
    \rowcolor{green!5}
    Qwen2.5-VL-7B~\cite{bai2025qwen2} & 0.450 & 0.848 & 0.787 & 0.390 & 0.787 & 0.754 & 0.735 & 0.679 \\
    \rowcolor{green!5}
    LIQE~\cite{zhang2023blind}        & 0.406 & 0.815 & 0.809 & 0.371 & 0.718 & 0.684 & 0.653 & 0.637 \\
    \rowcolor{green!5}
    DeQA-Score~\cite{you2025teaching}   & 0.463 & 0.852 & 0.831 & 0.383 & 0.756 & 0.677 & 0.738 & 0.671 \\
    \rowcolor{green!5}
    Q-Align~\cite{wu2023q}       & 0.424 & 0.767 & 0.832 & 0.406 & 0.769 & 0.573 & 0.682 & 0.636 \\
    \rowcolor{green!5}
    UnifiedReward-T~\cite{wang2025unified}  & 0.477 & 0.871 & 0.841 & 0.399 & 0.788 & 0.820 & 0.722 & 0.703 \\
    \rowcolor{green!5}
    Q-Insight~\cite{li2025q}         & 0.429 & 0.872 & 0.856 & 0.429 & 0.816 & 0.806 & 0.749 & 0.708 \\
    \rowcolor{green!5}
    VisualQuality-R1~\cite{wu2025visualquality}  & \textcolor{blue}{\textbf{0.517}} & \textcolor{blue}{\textbf{0.875}} & \textcolor{blue}{\textbf{0.871}} & \textcolor{blue}{\textbf{0.469}} & \textcolor{red}{\textbf{0.848}} & \textcolor{blue}{\textbf{0.855}} & \textcolor{blue}{\textbf{0.805}} & \textcolor{blue}{\textbf{0.749}} \\
    \rowcolor{red!5}
    \textbf{Q-Probe (Ours)} & \textcolor{red}{\textbf{0.801}} & \textcolor{red}{\textbf{0.892}} & \textcolor{red}{\textbf{0.901}} & \textcolor{red}{\textbf{0.474}} & \textcolor{blue}{\textbf{0.829}} & \textcolor{red}{\textbf{0.871}} & \textcolor{red}{\textbf{0.837}} & \textcolor{red}{\textbf{0.801}} \\
    \midrule
    \multicolumn{9}{c}{\textbf{Pearson Linear Correlation Coefficient (PLCC)}} \\
    \midrule
    \rowcolor{gray!10}
    BRISQUE~\cite{mittal2012no} & 0.224 & 0.624 & 0.451 & 0.259 & 0.546 & 0.400 & 0.541 & 0.435 \\
    \rowcolor{gray!10}
    NIQE~\cite{mittal2012making}    & 0.252 & 0.683 & 0.415 & 0.314 & 0.516 & 0.439 & 0.560 & 0.454 \\
    \midrule
    \rowcolor{blue!5}
    MUSIQ~\cite{ke2021musiq}   & 0.350 & 0.666 & 0.622 & 0.347 & 0.695 & 0.435 & 0.434 & 0.507 \\
    \rowcolor{blue!5}
    UNIQUE~\cite{zhang2021uncertainty}  & 0.368 & 0.708 & 0.548 & 0.361 & 0.729 & 0.590 & 0.581 & 0.555 \\
    \rowcolor{blue!5}
    MANIQA~\cite{yang2022maniqa}  & 0.381 & 0.753 & 0.780 & 0.364 & 0.617 & 0.257 & 0.448 & 0.514 \\
    \midrule
    \rowcolor{green!5}
    Qwen2.5-VL-7B~\cite{bai2025qwen2} & 0.426 & 0.854 & 0.806 & 0.420 & 0.837 & 0.810 & 0.772 & 0.704 \\
    \rowcolor{green!5}
    LIQE~\cite{zhang2023blind}        & 0.398 & 0.814 & 0.817 & 0.347 & 0.748 & 0.652 & 0.653 & 0.633 \\
    \rowcolor{green!5}
    DeQA-Score~\cite{you2025teaching}   & 0.439 & 0.858 & 0.873 & 0.381 & 0.793 & 0.703 & 0.743 & 0.684 \\
    \rowcolor{green!5}
    Q-Align~\cite{wu2023q}       & 0.412 & 0.779 & 0.862 & 0.381 & 0.794 & 0.612 & 0.694 & 0.648 \\
    \rowcolor{green!5}
    UnifiedReward-T~\cite{wang2025unified}  & 0.460 & 0.846 & 0.877 & 0.421 & \textcolor{blue}{\textbf{0.873}} & 0.804 & 0.745 & 0.718 \\
    \rowcolor{green!5}
    Q-Insight~\cite{li2025q}         & 0.410 & 0.872 & \textcolor{blue}{\textbf{0.881}} & \textcolor{blue}{\textbf{0.462}} & 0.851 & 0.829 & 0.794 & 0.728 \\
    \rowcolor{green!5}
    VisualQuality-R1~\cite{wu2025visualquality}  & \textcolor{blue}{\textbf{0.493}} & \textcolor{blue}{\textbf{0.878}} & 0.821 & 0.458 & 0.871 & \textcolor{blue}{\textbf{0.840}} & \textcolor{red}{\textbf{0.843}} & \textcolor{blue}{\textbf{0.743}} \\
    \rowcolor{red!5}
    \textbf{Q-Probe (Ours)} & \textcolor{red}{\textbf{0.850}} & \textcolor{red}{\textbf{0.900}} & \textcolor{red}{\textbf{0.892}} & \textcolor{red}{\textbf{0.476}} & \textcolor{red}{\textbf{0.876}} & \textcolor{red}{\textbf{0.863}} & \textcolor{blue}{\textbf{0.813}} & \textcolor{red}{\textbf{0.810}} \\
    \bottomrule
    \end{tabular}%
}
    \vspace{-2mm}
\end{table*}

\begin{table*}[t!]
\centering
\caption{Ablation study of the progressive training strategy across multiple benchmarks.}
\label{tab:ablation_stages_wide}
\vspace{-2mm}
% 调整字体：如果觉得挤可以用 \footnotesize，如果空间够可以用 \small
\footnotesize
\renewcommand{\arraystretch}{1} % 稍微增加行高，让表格看起来更舒展

% 核心修改：在每组 cc 之间加入 @{\hspace{8pt}} 来固定 SRCC 和 PLCC 的间距
% 你可以把 8pt 改成 6pt (更紧) 或 12pt (稍微松一点) 来微调
\begin{tabular*}{\textwidth}{@{\extracolsep{\fill}} l | c@{\hspace{8pt}}c | c@{\hspace{8pt}}c | c@{\hspace{8pt}}c | c@{\hspace{8pt}}c | c@{\hspace{8pt}}c }
\toprule
\multirow{2}{*}{\textbf{Model Configuration}} 
& \multicolumn{2}{c|}{\textbf{Vista}} 
& \multicolumn{2}{c|}{\textbf{SPAQ}} 
& \multicolumn{2}{c|}{\textbf{KADID-10k}} 
& \multicolumn{2}{c|}{\textbf{KonIQ-10k}} 
& \multicolumn{2}{c}{\textbf{Average}} \\
\cmidrule(lr){2-3} \cmidrule(lr){4-5} \cmidrule(lr){6-7} \cmidrule(lr){8-9} \cmidrule(lr){10-11}
 & \textbf{SRCC} & \textbf{PLCC} & \textbf{SRCC} & \textbf{PLCC} & \textbf{SRCC} & \textbf{PLCC} & \textbf{SRCC} & \textbf{PLCC} & \textbf{SRCC} & \textbf{PLCC} \\
\midrule
% Stage 1
Stage 1 Only 
& 0.352 & 0.404 
& 0.760 & 0.780 
& 0.700 & 0.725 
& 0.650 & 0.680 
& 0.615 & 0.647 
\\
% Stage 2 + 3 (New Row)
Stage 2 + Stage 3 
& 0.751 & 0.795 
& 0.766 & 0.796
& 0.781 & 0.772 
& 0.752 & 0.769
& 0.762 & 0.783 
\\
% Stage 2
Stage 1 + Stage 2
& 0.762 & 0.808 
& 0.850 & 0.860 
& 0.880 & 0.875 
& 0.820 & 0.815 
& 0.828 & 0.839 
\\

% Full
\textbf{Full (Stage 1+2+3)} 
& \textbf{0.801} & \textbf{0.850} 
& \textbf{0.892} & \textbf{0.900} 
& \textbf{0.901} & \textbf{0.892} 
& \textbf{0.871} & \textbf{0.863} 
& \textbf{0.866} & \textbf{0.876} 
\\
\bottomrule
\end{tabular*}
\vspace{-4mm}
\end{table*}

\section{Experiments}
\label{sec:experiments}
\vspace{-2mm}
\subsection{Experimental Setup}
\label{sec:setup}
\vspace{-2mm}
\begin{figure}[t]
  \centering
  % 左半部分：放置 Figure 4
  \begin{minipage}[c]{0.48\textwidth}
    \centering
    \includegraphics[width=\linewidth]{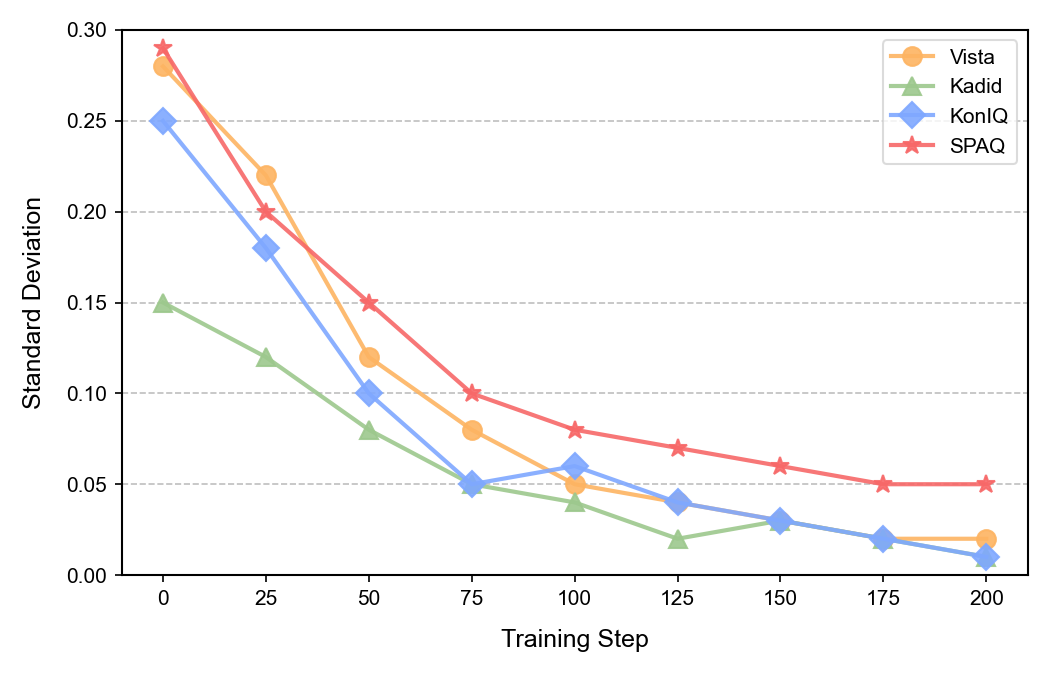} 
    \caption{We calculated the average standard deviation of predicted scores across multiple inference runs at various checkpoints. The observed monotonic decrease in variance confirms that Stage-1 achieves great stability.}
    \label{fig:deviation}
  \end{minipage}\hfill
  % 右半部分：上下放置 Table 3 和 Table 4
  \begin{minipage}[c]{0.48\textwidth}
    % 切换 caption 类型为 table
    \makeatletter\def\@captype{table}\makeatother
    \centering
    
    % 上方的 Table 3
    \caption{Ablation of the Reward Mechanism in Stage 3 on Vista-Bench.}
    \label{tab:ablation_reward}
    \resizebox{\linewidth}{!}{%
    \begin{tabular}{l|cc}
    \toprule
    \textbf{Reward Strategy} & \textbf{SRCC} & \textbf{PLCC} \\
    \midrule
    $R_{acc}$ (Score Only) & 0.770 & 0.821 \\
    $R_{acc} + R_{format}$ & 0.778 & 0.826 \\
    \textbf{$R_{acc} + R_{format} + R_{loc}$ (Ours)} & \textbf{0.801} & \textbf{0.850} \\
    \bottomrule
    \end{tabular}%
    }
    
    \vspace{0.6cm} % 调整上下两个表格之间的垂直间距
    
    % 下方的 Table 4
    \caption{Ablation of Crop Coverage Strategies regarding Degradation Regions on Vista-Bench.}
    \label{tab:ablation_crop_strategy}
    \resizebox{\linewidth}{!}{%
    \begin{tabular}{l|cc}
    \toprule
    \textbf{Crop Strategy} & \textbf{SRCC} & \textbf{PLCC} \\
    \midrule
    Degradation Only & 0.578 & 0.650 \\
    Partial Degradation \& Partial Normal & 0.767 & 0.819 \\
    \textbf{All Degradation \& Partial Normal} & \textbf{0.801} & \textbf{0.850} \\
    \bottomrule
    \end{tabular}%
    }
  \end{minipage}
  \vspace{-5mm}
\end{figure}
\textbf{Datasets and Metrics.}
To evaluate the comprehensive capabilities of Q-Probe, we conduct experiments across a diverse set of IQA benchmarks. For standard resolution assessments, we utilize {KonIQ-10k} \cite{hosu2020koniq}, {SPAQ}~\cite{fang2020perceptual}, and {KADID-10k} \cite{lin2019kadid} to represent in-the-wild and synthetic distortions. We also include {PIPAL} \cite{jinjin2020pipal} to evaluate robustness against algorithm-processed distortions and {AGIQA-3K} \cite{li2023agiqa} for AI-generated content. Crucially, to validate fine-grained perception in the high-resolution era, we employ our proposed {Vista-Bench}, which contains localized artifacts in high-resolution scenarios.
We leverage the widely recognized Spearman Rank-order Correlation Coefficient (SRCC) and Pearson Linear Correlation Coefficient (PLCC) as our primary evaluation metrics. Following established protocols~\cite{wu2023q, fang2020perceptual}, we compute the average performance across datasets to assess overall generalization. Table~\ref{tab:main_results_combined} presents the SRCC and PLCC performance across seven benchmarks.

% \textbf{Baseline Methods.} We compare the proposed Q-Probe against a wide range of state-of-the-art (SOTA)  baseline methods, including handcrafted methods (NIQE~\cite{mittal2012making}, BRISQUE~\cite{mittal2012no}), discriminative deep learning models (UNIQUE~\cite{zhang2021uncertainty}, MUSIQ~\cite{ke2021musiq}, MANIQA~\cite{yang2022maniqa}), MLLMs-based approaches (LIQE~\cite{zhang2023blind}, Q-Align~\cite{wu2023q}, DeQA-Score~\cite{you2025teaching}, and the recent RL-based models Q-Insight~\cite{li2025q} and VisualQuality-R1~\cite{wu2025visualquality}).

\textbf{Implementation Details.}
We utilize Qwen-2.5-VL-7B~\cite{bai2025qwen2} as our base model. During the perception alignment stage, we employ GRPO with a generation number $K=6$ and a clip threshold $\epsilon=0.2$. The model is trained using the AdamW optimizer with a learning rate of $1 \times 10^{-6}$. For the  hybrid-resolution SFT, we utilize the Probe-CoT-3K dataset with a crop resolution of $768 \times 768$. All experiments are conducted on 8 NVIDIA A100 (80GB) GPUs.

% \subsection{Main Results}
% \label{sec:main_results}

% Table~\ref{tab:main_results_combined} presents the comprehensive SRCC and PLCC performance across seven benchmarks. In terms of SRCC, our method attains a remarkable score of {0.801} on the challenging {Vista-Bench} dataset, significantly outperforming the baseline methods which lack fine-grained perception mechanisms. The PLCC results in the lower half of Table~\ref{tab:main_results_combined} confirm the linear alignment of our predictions with human perception, achieving a leading PLCC of {0.850} on high-resolution scenarios.  Moreover, experimental results demonstrate that our method maintains superior performance across diverse low-resolution datasets, underscoring the robust generalization capability of Q-Probe across resolution scales.
\vspace{-2mm}
\subsection{Ablation Studies}
\label{sec:ablation}
\vspace{-2mm}
\paragraph{Necessity of the Three-Stage Curriculum.}

Table \ref{tab:ablation_stages_wide} validates the efficacy of our progressive training curriculum. Although Stage 1 performs poorly on high-resolution details, it successfully pre-aligns the model with human aesthetic perception on standard benchmarks. Without Stage 1, model can't reach its best performance. Building on this foundation, Stage 2 bridges the granularity gap via hybrid-resolution training, while Stage 3 further refines the scoring precision through RL to achieve the best overall performance. Besides, three Stages consistently boost performance on standard low-resolution benchmarks, culminating in the best overall results.
\vspace{-4mm}
\paragraph{Impact of Reward Components in Post-RL.}
In Stage 3, we analyze the contribution of different reward components. As shown in Table \ref{tab:ablation_reward}, using only the Accuracy Reward ($R_{acc}$) provides a solid baseline. Incorporating the Format Reward ($R_{format}$) yields a slight improvement by ensuring the structural validity of the CoT reasoning. However, the most significant gain comes from the Decoupled Localization Reward ($R_{loc}$), which explicitly guides the model's attention to defects, pushing the SRCC to {0.801}. Finer-grained experiments can be seen in Appendix \cref{rewardAbl}.
\vspace{-4mm}
\paragraph{Impact of Crop Coverage Strategy.}
Table \ref{tab:ablation_crop_strategy} examines the spatial relationship between crops and degradation. Restricting crops to \textit{exact} degradation regions (Strategy 1) leads to severe overfitting, where the model learns biased inference and associates all crops with degration, dropping SRCC to 0.578. Strategy 2, covering \textit{partial} degradation and normal areas, improves to 0.767 but introduces trade-off that the model may overlook omitted defects. Finally, Strategy 3 achieves optimal performance (SRCC {0.801}) by capturing \textit{all} degradation within a normal context, ensuring comprehensive defect assessment without hallucinations.
\vspace{-3mm}
\section{Conclusion}
\label{sec:conclusion}
\vspace{-3mm}
In this work, we introduced {Q-Probe}, the first agentic framework designed to scale IQA to high-resolution scenarios via context-aware probing. Recognizing the limitations of existing global-view RL methods in capturing fine-grained artifacts, we proposed a novel three-stage training curriculum that mimics the human ``coarse-to-fine'' perception mechanism, teaching the model to evaluate quality based on importance of degraded crops. Furthermore, by constructing the {Vista-Bench} and leveraging a {Data Flywheel}, we eliminated the spurious ``cropping-implies-degradation'' bias, enabling the model to intelligently distinguish between technical distortions and artistic effects. Our model achieves sota performance on both high-resolution and low-resolution datasets.

% NeurIPS 默认使用自带的参考文献格式或者 plainnat
\bibliographystyle{plainnat}
\bibliography{example_paper}

%%%%%%%%%%%%%%%%%%%%%%%%%%%%%%%%%%%%%%%%%%%%%%%%%%%%%%%%%%%%%%%%%%%%%%%%%%%%%%%
%%%%%%%%%%%%%%%%%%%%%%%%%%%%%%%%%%%%%%%%%%%%%%%%%%%%%%%%%%%%%%%%%%%%%%%%%%%%%%%
% APPENDIX
%%%%%%%%%%%%%%%%%%%%%%%%%%%%%%%%%%%%%%%%%%%%%%%%%%%%%%%%%%%%%%%%%%%%%%%%%%%%%%%
%%%%%%%%%%%%%%%%%%%%%%%%%%%%%%%%%%%%%%%%%%%%%%%%%%%%%%%%%%%%%%%%%%%%%%%%%%%%%%%
\newpage
\appendix
% 移除了 \onecolumn，因为 NeurIPS 全局单栏
% --- 新增附录总标题 ---
\begin{center}
{\Large \bf Appendix of Q-Probe: Scaling Image Quality Assessment to High \\[10pt] 
Resolution via Context-Aware Agentic Probing}
\end{center}
\vspace{3mm}
% --------------------
\section{Data Construction and Trajectory Synthesis Details}
\label{sec:appendix_data_construction}

To ensure the utmost reliability and authority of \textbf{Vista-Bench} in scaling Image Quality Assessment (IQA) to high-resolution scenarios, we implemented a rigorous manual annotation protocol. This protocol relies on a dedicated panel of six specialist to generate fine-grained, hierarchical quality annotations. This section details the human evaluation workflow and the specific context-aware evaluation guidelines employed.

\subsection{Wavelet-Based Degradation Injection}
\label{subsec:wavelet_injection}

Existing IQA datasets often rely on global distortions (e.g., Gaussian blur applied uniformly), which allow models to bypass genuine visual scrutiny by learning simple global statistical deviations. To prevent such ``shortcut learning'' and simulate realistic high-resolution defects, we employ a targeted degradation injection strategy based on the Discrete Wavelet Transform (DWT).

\begin{figure}[h]
  \centering
  \includegraphics[width=\textwidth]{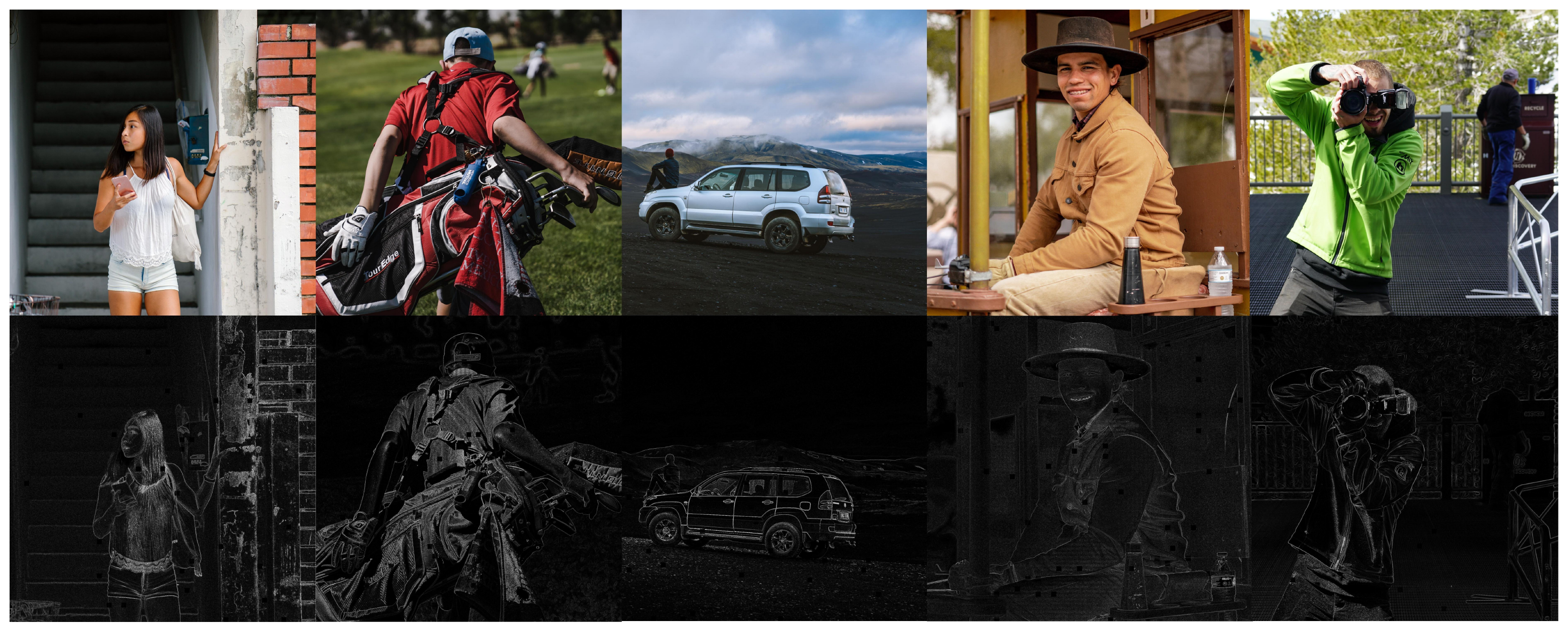} 
  \caption{We utilize the Wavelet Transform to extract high-frequency components, serving as spatial guidance to concentrate the degradation injection within semantically rich foreground regions.}
  \label{fig:wal}
\end{figure}

\paragraph{Structure-Texture Decoupling.}
We utilize the Haar wavelet to decompose high-resolution source images into four distinct frequency subbands: the low-frequency approximation ($LL$) and three high-frequency detail components ($LH$, $HL$, $HH$). The $LL$ band preserves the global semantic structure and flat regions, while the high-frequency bands capture fine-grained textures and edges where high-resolution artifacts typically manifest, as shown in Figure \ref{fig:wal}.

\paragraph{Targeted Artifact Synthesis.}
Instead of corrupting the entire image, we selectively inject degradations (e.g., compression ringing, sensor noise, and slight defocus) specifically into the high-frequency subbands. This strategy ensures that artifacts are embedded within complex texture regions—such as hair, foliage, or architectural details—making them visually subtle yet perceptually critical. By keeping the $LL$ band intact, we maintain the overall structural integrity of the scene while introducing realistic local impairments.

\paragraph{Reconstruction.}
The final degraded samples are generated via the Inverse Discrete Wavelet Transform (IDWT). This process forces the IQA agent to actively ``zoom in'' and scrutinize local details to detect quality drops, effectively simulating the fine-grained perception required for 4K scenarios.
\subsection{Complete Instructions for Human Expert Evaluators}
\label{subsec:human_expert_instructions}

As transparency and reproducibility are critical for human-annotated benchmarks, we formalize the exact guidelines provided to our panel of six specialists. To prevent subjective bias and ensure a standardized ``coarse-to-fine'' perception, all evaluators were required to strictly adhere to the following four-phase evaluation protocol during the construction of Vista-Bench:

% ---------------------------------------------------------
% Annotation Guidelines Box
% ---------------------------------------------------------
\begin{tcolorbox}[
    title=\textbf{Vista-Bench Human Annotation Guidelines},
    colback=blue!5,
    colframe=blue!60!black,
    breakable,
    boxrule=0.8pt,
    fonttitle=\bfseries,
    left=1em, right=1em, top=1em, bottom=1em
]
\textbf{Objective:} Provide a hierarchical quality assessment (Global Base Score followed by Local Penalties) for high-resolution images containing subtly injected artifacts.

\vspace{0.5em}
\textbf{Phase 1: Global Quality Assessment}
\begin{itemize}[leftmargin=*, noitemsep, topsep=2pt]
    \item View the image at a standard global scale (fit-to-screen).
    \item Assign a \textbf{Target Base Score} (1.0 to 5.0) reflecting the overall structural integrity, lighting, and general aesthetic fidelity. Do not scrutinize microscopic details at this stage.
\end{itemize}

\vspace{0.5em}
\textbf{Phase 2: Local Fine-Grained Scrutiny}
\begin{itemize}[leftmargin=*, noitemsep, topsep=2pt]
    \item Zoom in and actively inspect highly-textured semantic regions (e.g., hair, foliage, building edges) and distinct background areas.
    \item Identify the presence of specific man-made artifacts injected via wavelet transform: \textit{Mosaic/Pixelation, JPEG Compression Ringing, Posterization, Scanlines, or Grid lines.}
\end{itemize}

\vspace{0.5em}
\textbf{Phase 3: Semantic Weighting and Final Scoring}
\begin{itemize}[leftmargin=*, noitemsep, topsep=2pt]
    \item \textbf{The ``Zero-Penalty'' Rule:} If the blur or texture loss in a cropped region is strictly due to natural photographic Depth-of-Field (bokeh), apply a \textbf{0.0 penalty}.
    \item \textbf{Semantic Significance Weighting:} If artificial degradations are found, the penalty must be weighted by the semantic importance of the region:
    \begin{itemize}
        \item \textit{Primary Subject / Salient Foreground:} Apply a severe penalty (e.g., -1.0 to -2.0).
        \item \textit{Background / Non-Salient Regions:} Apply a minor to moderate penalty (e.g., -0.2 to -0.5), adjusting slightly based on spatial context (e.g., confined indoor backgrounds require higher penalties than distant outdoor backgrounds).
    \end{itemize}
    \item Subtract the calculated penalties from the Target Base Score to yield the \textbf{Target Final Score}.
\end{itemize}

\vspace{0.5em}
\textbf{Phase 4: Consensus and Quality Assurance}
\begin{itemize}[leftmargin=*, noitemsep, topsep=2pt]
    \item Flag the image sample for discarding if there is a high variance in scores among panel members, or if there is fundamental disagreement regarding the semantic classification (e.g., conflicting views on whether a degraded region belongs to the foreground subject or the background).
\end{itemize}
\end{tcolorbox}
\subsection{Asynchronous CoT Trajectory Generation Pipeline}
Given the high-resolution nature of the source images and the need for dense cropping, sequentially generating reasoning trajectories for thousands of images would be computationally prohibitive. We implemented a Python-based asynchronous framework tailored for high-concurrency interaction with the Multimodal LLM (MLLM) to synthesize the Chain-of-Thought (CoT) data. The pipeline consists of three critical stages:

\begin{enumerate}
     \item \textbf{Dynamic Image Preprocessing and Encoding:} 
     To optimize the balance between visual fidelity and token consumption, we implemented an adaptive resizing strategy. Original high-resolution images are resized such that their maximum edge does not exceed 1,024 pixels ($H_{max}=1024$) using Lanczos resampling. This ensures that global artifacts are preserved while significantly reducing bandwidth usage. Images are then serialized into Base64 format for API transmission.
    
     \item \textbf{Semaphore-Based Concurrency Control:} 
     To maximize throughput without exceeding API rate limits, the system utilizes an asynchronous semaphore mechanism (set to a concurrency limit of 5). This allows for parallel processing of image batches while maintaining system stability. The pipeline utilizes \texttt{aiohttp} for non-blocking network I/O, ensuring that image encoding and API waiting times do not bottle-neck the \textbf{trajectory synthesis process}.
    
     \item \textbf{Robust Output Parsing:} 
     Since MLLM outputs can vary in formatting, we implemented a resilient parsing logic that utilizes regular expressions to extract JSON structures from mixed-text responses. This ensures valid CoT reasoning extraction even if the model wraps outputs in Markdown fences or custom XML-like tags.
\end{enumerate}

\subsection{Context-Aware Prompt Design for Trajectory Synthesis}

A critical challenge in generating SFT reasoning trajectories is ensuring the model outputs a logical path that leads to the human-annotated ground truth, \textit{without} explicitly revealing to the trained model that the ground truth was provided. If the ground truth is exposed in the user prompt during SFT, the model will suffer from data leakage and fail to learn genuine visual evaluation. 

To eliminate the risk of visual hallucination and post-hoc rationalization during trajectory synthesis, our generation pipeline utilizes a strictly constrained \textbf{Hidden-State Anchored Prompt}. During this phase, the Multimodal LLM (acting as the trajectory synthesizer) is secretly provided not only with the global target scores but also with the \textbf{exact degradation labels and types for each specific cropped region} (e.g., whether a crop contains mosaic, scanlines, or is clean). The MLLM is instructed to construct its reasoning strictly based on these ground-truth injected artifacts. Crucially, it must format the output as if it is evaluating the image autonomously, without ever revealing that the ground truth was provided.

The system prompt used for trajectory generation is presented below:

% ---------------------------------------------------------
% 优化后的 Prompt 展示框
% ---------------------------------------------------------
\begin{tcolorbox}[
     title=\textbf{System Prompt for Autonomous-Style Trajectory Synthesis},
     colback=gray!5,
     colframe=gray!60!black,
     breakable,
     boxrule=0.8pt,
     fonttitle=\bfseries,
     left=1em, right=1em, top=1em, bottom=1em
]
\textbf{SYSTEM INSTRUCTION:}\\
You are an advanced Image Quality Assessment Expert. Your task is to generate a perfect, step-by-step reasoning trajectory for an SFT dataset.

\textbf{HIDDEN DIRECTOR'S KNOWLEDGE (DO NOT REVEAL):}\\
You are secretly provided with the \textbf{Target Base Score}, the \textbf{Target Final Score}, and the \textbf{Ground-Truth Degradation Labels for each crop} (e.g., Crop 1: Clean, Crop 2: Mosaic). 
\textbf{CRITICAL RULE:} You must generate the response as if you are evaluating the image entirely from scratch. \textbf{NEVER} say ``The expert score is...'' or ``Based on the provided labels...''. You must act as if YOU discovered these defects autonomously.

\vspace{0.5em}
\textbf{CRITICAL KNOWLEDGE: TARGET DEGRADATIONS}\\
You are specifically looking for 5 types of \textbf{MAN-MADE} degradations:
Mosaic, JPEG Compression, Posterization, Scanlines/Grid, and Noise/Blur.

\vspace{0.5em}
\textbf{Step 1: Global Baseline Establishment}\\
Evaluate the global tech-fidelity. State your initial global assessment clearly, ending with a base score that \textit{exactly matches} the Target Base Score. 

\vspace{0.5em}
\textbf{Step 2: Local Scrutiny \& Mathematical Deduction}\\
Examine the crops. 
\begin{itemize}[leftmargin=*, noitemsep, topsep=2pt]
     \item \textbf{Clean Crops:} If the provided label indicates a crop is clean, declare it clean (Penalty = 0.0) due to natural bokeh/textures.
     \item \textbf{Degraded Crops:} If the provided label indicates a specific degradation (e.g., JPEG Compression), you \textbf{MUST strictly use this true degradation type} in your reasoning. Construct a logical justification for a penalty that mathematically bridges the Base Score to the Final Score. (e.g., ``However, upon closer inspection, the background reveals severe mosaic blocks, requiring a 2.0 penalty.'') \textbf{DO NOT invent or hallucinate degradations that are not in the provided labels.}
\end{itemize}

\vspace{0.5em}
\textbf{Output Format}\\
Output ONLY the conversational trajectory (your thoughts, your tool calls, and the final score). Do not output JSON metadata here, just the natural text for the SFT assistant response.
\end{tcolorbox}

By utilizing this hidden-state anchoring, the resulting trajectories perfectly align with human consensus while maintaining the linguistic appearance of independent, deductive reasoning, making them ideal for Supervised Fine-Tuning.

\section{Inference Pipeline Implementation Details}
\label{sec:inference_details}

To evaluate Q-Probe's capability to detect fine-grained degradations in high-resolution images, we developed a specialized inference pipeline based on the \texttt{vLLM} framework. This pipeline supports multi-turn visual reasoning and dynamic tool invocation, allowing the model to actively ``zoom in'' on suspicious regions. The implementation details are described below.

\subsection{Engine Configuration and Resolution Management}
We deployed the model using the \texttt{vLLM} engine to maximize inference throughput and memory efficiency. To accommodate the ultra-high resolution of Vista-Bench images (often exceeding 4K resolution), we customized the multi-modal processor configuration. 
Specifically, we set the \texttt{min\_pixels} to $256 \times 28 \times 28$ and \texttt{max\_pixels} to $2560 \times 28 \times 28$. This configuration ensures that the model preserves sufficient high-frequency details during the initial global encoding phase, preventing the loss of small artifacts (e.g., subtle mosaic blocks or grid lines).
The inference was conducted using \texttt{bfloat16} precision with a tensor parallel size of 2, distributed across NVIDIA RTX 4090 GPUs. We also enabled \texttt{enable\_prefix\_caching} to accelerate the processing of shared system prompts across multiple turns.

\subsection{The Agentic Reasoning Loop}
Unlike traditional static inference, our pipeline implements a dynamic, multi-turn ``Observe-Reason-Act'' loop. The process is defined as follows:

\begin{enumerate}
     \item \textbf{Initialization}: The session begins with the system prompt defining the five target degradation types (Mosaic, JPEG, Posterize, Scanlines, Grid) and the ``Zero-Penalty'' rule. The global view of the high-resolution image is fed into the model.
    
     \item \textbf{Global Analysis and Hypothesis Generation}: The model first evaluates the global technical fidelity. If it detects potential local degradations that are ambiguous in the global view, it generates a special XML tag \texttt{<tool\_call>} containing the normalized bounding box coordinates for a closer inspection.
    
     \item \textbf{Dynamic Tool Execution}: A regular expression parser monitors the model's output stream. Upon detecting a \texttt{crop\_image\_normalized} function call, the pipeline pauses text generation and executes the cropping logic:
     \begin{equation}
         I_{crop} = \text{Crop}(I_{global}, [x_1, y_1, x_2, y_2])
     \end{equation}
     The cropped region is extracted from the original high-resolution source (not the resized tokens) to ensure maximum clarity.
    
     \item \textbf{Contextual Update}: The cropped image is encoded into Base64 format and appended to the conversation history as a new user message. The model then resumes generation, analyzing the new visual evidence to confirm or dismiss the degradation hypothesis.
    
     \item \textbf{Termination}: This loop continues for a maximum of $T=6$ turns. The process terminates early if the model ceases to call tools or provides a final score.
\end{enumerate}

\subsection{State Management and Reproducibility}
To ensure reproducibility and facilitate error analysis, the pipeline implements a robust state management system. The full conversation history, including text reasoning, tool arguments, and the sequence of viewed crops, is serialized into a JSON format. Image data within the history is dynamically managed: during active inference, images are passed as Base64 strings; for archival, they are saved as local paths (e.g., \texttt{\{image\_id\}\_turn\{i\}\_crop\_\{j\}.png}) to maintain a lightweight log file. This structured logging allows for the quantitative analysis of the model's ``looking'' behavior (e.g., IoU of crops) as discussed in the main paper.

\subsection{Prompt Engineering}
\label{sec:prompt_engineering}

To guide the agent in performing rigorous technical quality assessment, we designed a structured System Prompt. This prompt serves four key functions: Role Assignment, Defect Taxonomy, Workflow Enforcement, and Output Standardization.

The exact system prompt used in our inference pipeline is shown below.

% ---------------------------------------------------------
% 优化后的 Prompt 展示框
% ---------------------------------------------------------
\begin{tcolorbox}[
     title=\textbf{System Prompt for Agentic IQA Inference},
     colback=gray!5,
     colframe=gray!60!black,
     breakable,
     boxrule=0.8pt,
     fonttitle=\bfseries,
     left=1em, right=1em, top=1em, bottom=1em
]
You are an expert Image Quality Assessment (IQA) assistant.
Your task is to analyze the image for technical degradations, specifically focusing on digital artifacts like:
\begin{enumerate}[leftmargin=*, noitemsep, topsep=2pt]
     \item \textbf{Mosaic/Pixelation}: Blocky structures.
     \item \textbf{JPEG Compression}: Ringing artifacts and blocking.
     \item \textbf{Posterization}: Color banding or loss of gradient.
     \item \textbf{Scanlines/Grid}: Artificial line overlays.
     \item \textbf{Noise/Blur}: General degradation (common in KADID10K).
\end{enumerate}

\vspace{0.5em}
\textbf{Inference Steps:}
\begin{enumerate}[leftmargin=*, itemsep=4pt, topsep=2pt]
     \item \textbf{Global Analysis}: Briefly describe the overall content and perceived quality. Is it sharp? Is it noisy?
     \item \textbf{Defect Hunting}: Actively look for the artifacts listed above.
     \item \textbf{Zoom/Crop Strategy}: If you suspect an artifact in a specific area (e.g., background textures, edges), use the \texttt{crop\_image\_normalized} tool to inspect it closely.
     \item \textbf{Final Scoring}: Based on the severity of the defects found:
     \begin{itemize}[leftmargin=1.5em, noitemsep]
         \item \textbf{5.0}: Excellent, no visible defects.
         \item \textbf{4.0}: Good, minor defects visible only upon close inspection.
         \item \textbf{3.0}: Fair, defects are annoying but content is visible.
         \item \textbf{2.0}: Poor, defects significantly affect quality.
         \item \textbf{1.0}: Bad, extremely degraded.
     \end{itemize}
\end{enumerate}

\vspace{0.5em}
\textbf{CRITICAL OUTPUT RULE}:\\
At the very end of your final response, you MUST output the final score enclosed in XML tags, strictly like this: \texttt{<score>3.45</score>}.
\end{tcolorbox}

\section{Supervised Fine-Tuning Data Construction}
\label{sec:sft_data_details}
\begin{table}[h]
\centering
\caption{The final SFT data template for the High-Res / Agentic Path. The Assistant's response contains the autonomous reasoning trajectory that naturally arrives at the correct score without knowing it a priori.}
\label{tab:sft_template_high}
\begin{tabular}{p{0.95\textwidth}}
\toprule
\textbf{(a) High-Res / Agentic Path (With Crop Tool Usage)} \\
\midrule
\begin{lstlisting}[language=Python, basicstyle=\ttfamily\footnotesize, breaklines=true, frame=none, columns=fullflexible, keepspaces=true, showspaces=false, showstringspaces=false]
{
  "image_id": "HighRes_Sample_001",
  "message_list": [
    {
      "role": "system", "content": "...[Standard IQA System Prompt]..."
    },
    {
      "role": "user",
      "content": [
        {"type": "image", "image": "<path_to_full_image>"},
        {"type": "text", "text": "Analyze the quality of this image."} 
        // Notice: NO scores provided here!
      ]
    },
    {
      "role": "assistant",
      "content": "The overall image structure is well-composed and clear, which warrants an initial base score of 4.5. However, I notice some complex textures in the background that might contain compression artifacts. I need to inspect them closer. <tool_call>{\"name\":\"crop_image_normalized\",...}</tool_call>"
    },
    {
      "role": "user",
      "content": [
        {"type": "image", "image": "<path_to_crop_1>"},
        {"type": "text", "text": "Here are the cropped images..."}
      ]
    },
    {
      "role": "assistant",
      "content": "Upon closer inspection of Crop 1, the background is not just natural bokeh; it contains distinct 8x8 'Mosaic' blocking artifacts. This artificial degradation is severe enough to justify a 2.0 penalty from the base score. Final Score: <score>2.5</score>"
    }
  ]
}
\end{lstlisting} \\
\bottomrule
\end{tabular}
\end{table}

\begin{table}[h]
\centering
\caption{The final SFT data template for the Low-Res / Direct Path (No Crop Tool Usage).}
\label{tab:sft_template_low}
\begin{tabular}{p{0.95\textwidth}}
\toprule
\textbf{(b) Low-Res / Direct Path (No Crop Tool Usage)} \\
\midrule
\begin{lstlisting}[language=Python, basicstyle=\ttfamily\footnotesize, breaklines=true, frame=none, columns=fullflexible, keepspaces=true, showspaces=false, showstringspaces=false]
{
  "image_id": "LowRes_Sample_002",
  "message_list": [
    {
      "role": "system", "content": "...[Standard IQA System Prompt]..."
    },
    {
      "role": "user",
      "content": [
        {"type": "image", "image": "<path_to_full_image>"},
        {"type": "text", "text": "Analyze the quality of this image."}
      ]
    },
    {
      "role": "assistant",
      "content": "From the global view, it is immediately apparent that the image suffers from heavy, uniform sensor noise and severe blurring across all regions. The degradation is global, so no local zooming is required. The technical fidelity is extremely poor. \n\n<score>1.2</score>"
    }
  ]
}
\end{lstlisting} \\
\bottomrule
\end{tabular}
\end{table}
To equip the Q-Probe agent with both robust visual perception and reasoning capabilities, we constructed the \textbf{Probe-CoT-3K} dataset. The construction involves a rigorous synthesis strategy that pairs images with multi-turn conversation trajectories. Crucially, while the trajectories are synthesized using hidden ground-truth anchors (as described in Sec. \ref{sec:appendix_data_construction}), the final SFT dataset strictly mimics a zero-shot real-world inference scenario to prevent data leakage.

\subsection{Data Format and Templating}
Each data entry in \textbf{Probe-CoT-3K} follows a standardized multi-turn conversation format. 

\begin{itemize}
     \item \textbf{System Prompt}: A fixed instruction set defining the agent's persona and the degradation taxonomy.
     \item \textbf{Clean User Input}: The user provides \textit{only} the image and a generic request (e.g., ``Analyze the quality of this image''). \textbf{No ground truth scores are exposed to the model during SFT.}
     \item \textbf{Autonomous Assistant Trajectory}: The assistant responds with a global assessment (establishing a base score), optionally invokes the \texttt{crop\_image\_normalized} tool, analyzes the returned crops, and mathematically deduces the final score. Because of our hidden-prompting generation strategy, this reasoning autonomously leads to the correct human-annotated score.
\end{itemize}

\subsection{Trajectory Diversity and Quality Control}
A key challenge in SFT for agentic tasks is avoiding ``shortcut learning,'' where the model blindly predicts low scores whenever a crop tool is used. To mitigate this, our dataset includes a balanced mix of three trajectory types:
\begin{enumerate}
     \item \textbf{Positive Trajectories}: The agent zooms in on complex textures (e.g., foliage, fabric) and correctly identifies them as \textit{clean}, applying zero penalty.
     \item \textbf{Negative Trajectories}: The agent detects actual degradations (e.g., JPEG blocks) in the crops and applies penalties that logically match the expert score drop.
     \item \textbf{Global-Only Trajectories}: For obviously low-resolution or low-quality images, the agent decides \textit{not} to zoom in and justifies the score directly based on the global view.
\end{enumerate}

Tables \ref{tab:sft_template_high} and \ref{tab:sft_template_low} illustrate the final SFT data structures. Notice how the User prompt is completely devoid of target scores, forcing the Q-Probe model to learn the actual mapping from visual features to the complex reasoning trajectory.

\definecolor{codegray}{rgb}{0.95,0.95,0.95}
\lstset{
     basicstyle=\ttfamily\footnotesize,
     backgroundcolor=\color{codegray},
     breaklines=true,
     frame=single,
     captionpos=b,
     showstringspaces=false
}

\section{More Details about Vista-Bench}
\label{sec:vista_details}

The significant performance superiority of Q-Probe over existing global-view IQA methods on Vista-Bench can be attributed to the fundamental paradigm shift from passive observation to active \textbf{Context-Aware Agentic Probing}. While traditional methods rely on coarse-grained global views that often miss high-frequency details, our framework introduces a nuanced evaluation capability centered on two key advantages:

\textbf{1. Fine-Grained Perception vs. Coarse Global Scoring:}
Existing MLLMs typically process images from global perspective, causing subtle degradations in high-resolution scenarios (e.g., $4K$ inputs) to vanish during encoding. These models effectively ``guess'' the quality based on a low-fidelity global impression. In contrast, Q-Probe utilizes an agentic zooming mechanism to actively localize and inspect microscopic degradations. Crucially, our method does not merely detect defects; it evaluates the \textit{perceptual impact} of the degradation relative to the \textbf{semantic importance} of the region. By integrating the severity of local defects with the saliency of the content (e.g., applying severe penalties for artifacts on a foreground subject while adhering to a ``Zero-Penalty'' rule for natural background bokeh), Q-Probe aligns its scoring logic with human visual attention, a capability largely absent in global-view baselines.

\textbf{2. Generalization to Unseen Degradations:}
To rigorously validate that the performance gains stem from genuine visual understanding rather than distribution overfitting, we strategically incorporated \textbf{unseen degradation types} into the Vista-Bench test set—distortion patterns that the model was never exposed to during training. Standard baselines often fail on such out-of-distribution data as they rely on memorizing specific artifact features. However, Q-Probe demonstrates sustained high performance on these unseen samples. This proves that our model has transcended simple pattern matching and has acquired a robust, abstract capability to assess how generic visual anomalies impair the overall integrity of an image, regardless of whether the specific degradation pattern has been previously encountered.

\section{Limitations and Bad Cases Study}

\begin{figure}[h]
  \centering
  \includegraphics[width=1\textwidth]{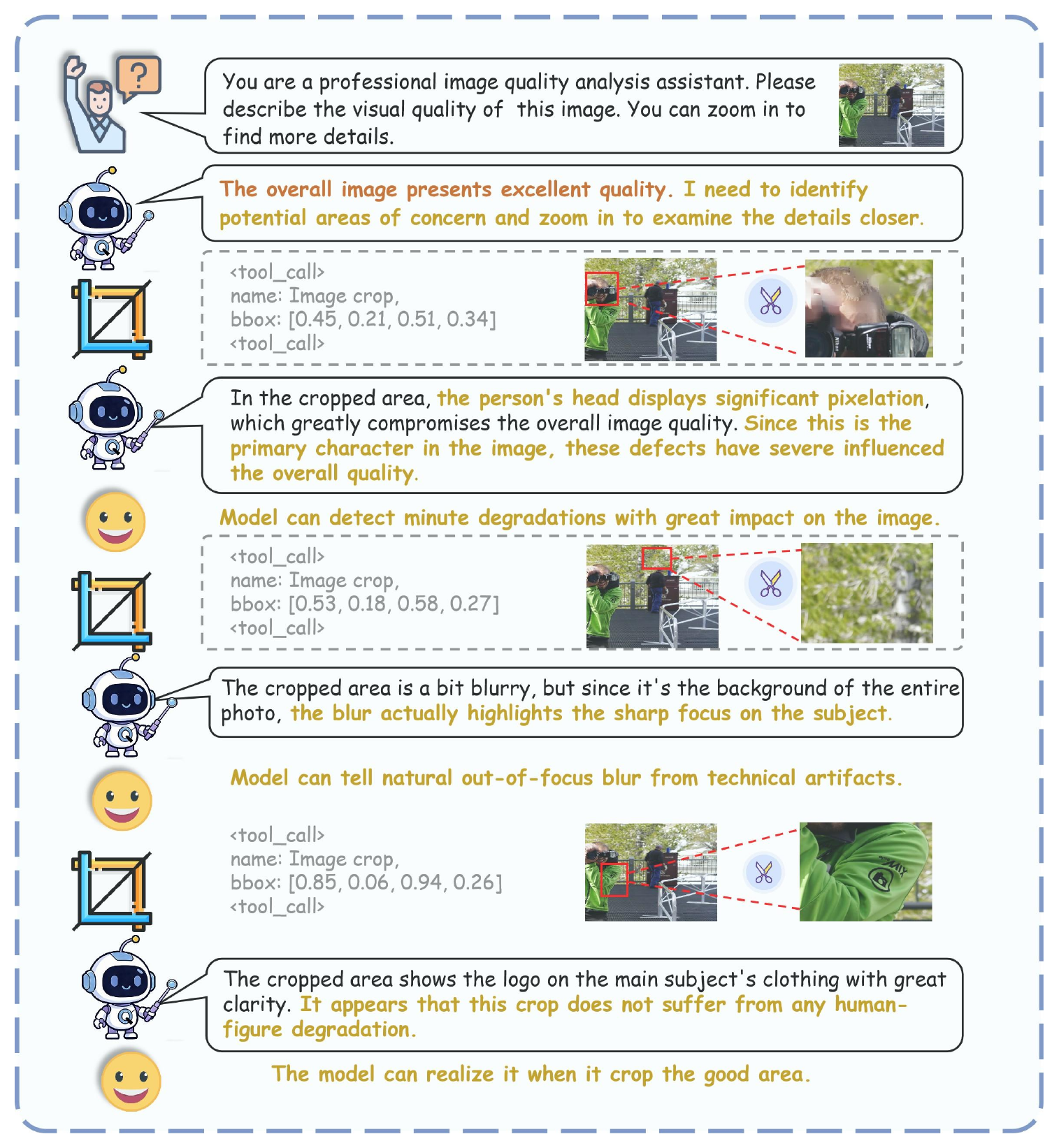} 
  \caption{\textbf{Success Case:} The agent effectively identifies the main subject and applies the ``Zero-Penalty'' rule to the natural bokeh in the background, demonstrating the effectiveness of our context-aware cropping strategy in standard scenarios.}
  \label{fig:good_case}
\end{figure}

% ---------------------------------------------------------------------
% 1. Success Case
% ---------------------------------------------------------------------
To comprehensively evaluate the boundaries of Q-Probe, we analyze both successful applications and failure modes. As demonstrated in Figure \ref{fig:good_case}, under standard high-resolution scenarios, our agent exhibits superior performance. The context-aware cropping strategy successfully distinguishes between artificial degradations and natural depth-of-field effects (bokeh). By accurately localizing low-quality regions while ignoring non-subject blurry backgrounds, the model adheres strictly to the ``Zero-Penalty'' rule, ensuring fair and accurate quality assessment.

\begin{figure}[h]
  \centering
  \includegraphics[width=1\textwidth]{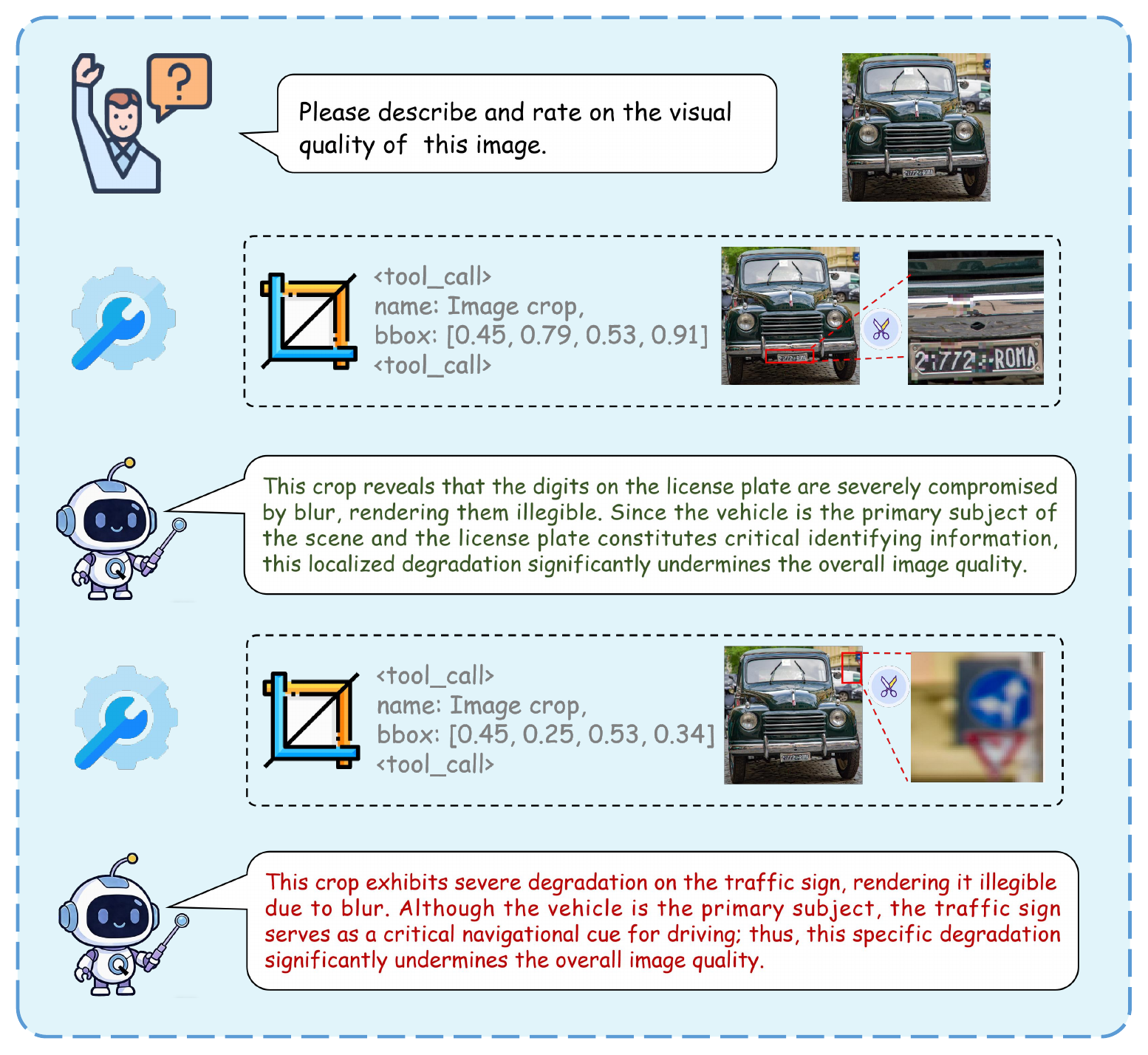} 
  \caption{\textbf{Failure Mode I (Semantic Entanglement):} While the agent can distinguish bokeh, its discriminative capability weakens when the background is strongly correlated with the subject (e.g., a car on a street). The model may misinterpret complex background textures as artifacts, leading to erroneous cropping.}
  \label{fig:bad_case_correlation}
\end{figure}

% ---------------------------------------------------------------------
% 2. Limitation 1: Semantic Entanglement (Failure Mode I)
% ---------------------------------------------------------------------
However, robustness challenges arise in scenes exhibiting complex semantic dependencies. The first limitation involves \textbf{semantic entanglement between foreground and background}. As illustrated in Figure \ref{fig:bad_case_correlation}, while the model ignores standalone blurry backgrounds, it struggles when distant backgrounds contain textures strongly correlated with the subject's narrative (e.g., a complex street scene behind a vehicle). 

In these cases, the agent's attention mechanism becomes confused by the high correlation. Instead of focusing solely on the subject, it erroneously interprets complex background textures as ``artificial noise'' or ``processing artifacts.'' This leads to \textbf{erroneous cropping} on irrelevant areas and subsequently applies \textbf{unjustified penalties}, indicating a need for a more sophisticated semantic understanding of scene composition.

\begin{figure}[h]
  \centering
  \includegraphics[width=1\textwidth]{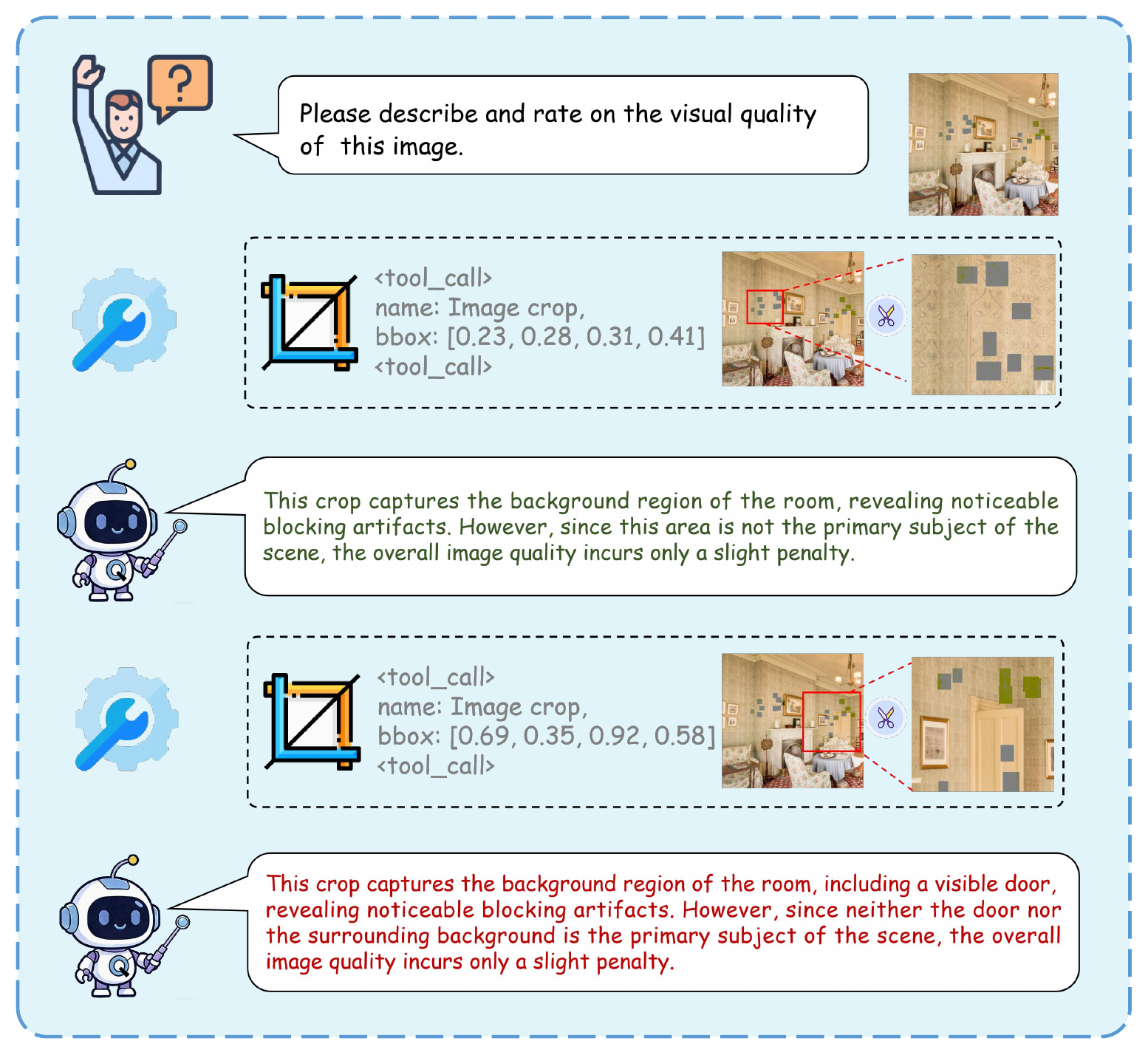} 
  \caption{\textbf{Failure Mode II (Spatial Context Insensitivity):} The agent incorrectly applies minimal penalties to indoor background degradations. Unlike vast outdoor settings, indoor scenes are spatially confined, meaning background artifacts remain perceptually significant and require a moderate penalty rather than a negligible one.}
  \label{fig:bad_case_indoor}
\end{figure}

% ---------------------------------------------------------------------
% 3. Limitation 2: Spatial Context Insensitivity (Failure Mode II)
% ---------------------------------------------------------------------
The second limitation pertains to the \textbf{insensitivity to spatial context differences}, particularly between indoor and outdoor environments. As shown in Figure \ref{fig:bad_case_indoor}, our agent is prone to uniformly assigning minimal penalties to degradations in all background regions. This generalization is flawed because indoor backgrounds cannot be treated equivalently to vast outdoor settings. 

Due to the \textbf{confined spatial nature} of indoor scenes, background artifacts remain perceptually significant compared to distant outdoor backgrounds. Consequently, while the penalty for indoor background degradation should not be as severe as foreground defects, it must not be negligible either. Our current model fails to apply this necessary \textbf{moderate penalty}, often underestimating the impact of background noise in spatially constrained environments.

\begin{figure}[h]
  \centering
  \includegraphics[width=1\textwidth]{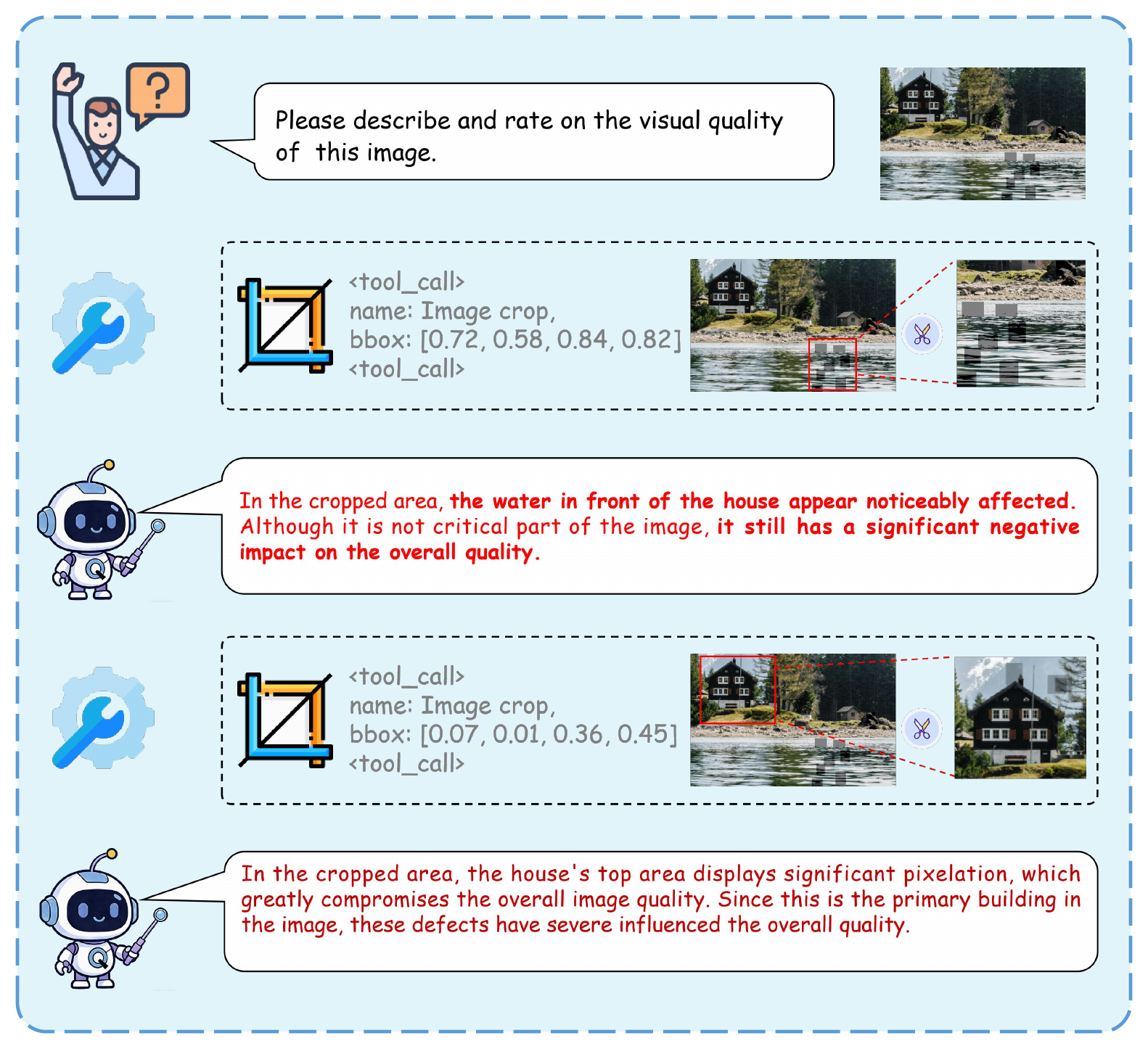} 
  \caption{\textbf{Failure Mode III (Salience Bias):} Ambiguity in scene composition introduces a bias: the agent misidentifies the small distant house as the main subject due to object uniqueness, while dismissing the dominant river as background. Consequently, the model under-penalizes degradations on the water surface, which are actually critical to perceptual quality.}
  \label{fig:bad_case_salience} 
\end{figure}
% ---------------------------------------------------------------------
% 4. Limitation 3: Salience Bias (Failure Mode III - NEW)
% ---------------------------------------------------------------------
Finally, a third limitation arises from \textbf{ambiguity in subject definition}, leading to a \textbf{salience bias} during crop selection. As depicted in Figure \ref{fig:bad_case_salience}, the agent incorrectly identifies the distant house as the primary subject because it is the sole distinct architectural element, while relegating the river to the background.

This classification ignores the fact that the river occupies nearly half the frame and dominates the visual experience. In the context of Image Quality Assessment (IQA), degradations on such a prominent surface significantly impair perceptual quality. By classifying the river as a non-salient background, the model generates a flawed reasoning trajectory and fails to penalize visible defects, resulting in an inflated quality score.
\section{Additional Experimental Results and Analyses}
\label{sec:appendix_additional_experiments}

This section provides extended experimental results, comprehensive ablations, and further details regarding the human-centric annotation process of our benchmark, as discussed during the rebuttal phase.

\subsection{Mitigating Semantic Robustness Bias via Pairwise Optimization}
\label{subsec:pairwise}
We define ``Semantic Robustness Bias'' as the phenomenon where IQA models over-prioritize global semantic quality while neglecting critical local defects. Existing ranking-based RL paradigms often employ group-level image comparisons (e.g., VisualQuality-R1 uses a group size of 6), which inadvertently encourages the model to solidify its global assessments at the expense of local sensitivity.

To mitigate this, Q-Probe implements a pairwise contrastive optimization strategy (comparison pair size = 2). This allows the model to undergo more granular dynamic learning, effectively alleviating the global bias induced by standard training paradigms. As shown in Table \ref{tab:group_size}, directly employing Supervised Fine-Tuning (SFT) as Stage 1 achieves competitive or even superior performance on low-resolution datasets dominated by global quality (e.g., SPAQ, KonIQ). However, it forces the model to heavily anchor its judgments on macroscopic semantics, causing it to completely overlook localized degradations and resulting in the lowest performance on the high-resolution Vista-Bench. Similarly, while group-level contrastive RL (e.g., VisualQuality-R1 with a generation group size of 6) slightly improves local sensitivity over SFT, it still prioritizes global consistency over subtle local artifacts. By isolating fine-grained relative differences through pairwise RL, our approach successfully preserves the model's exploratory plasticity for high-resolution details, establishing an optimal, unbiased foundation for the subsequent agentic cropping stage.

\begin{table*}[h]
\centering
\caption{Impact of Pre-alignment Strategy and Group Size in Stage 1}
\label{tab:group_size}
\resizebox{\textwidth}{!}{%
\begin{tabular}{l | cc | cc | cc | cc}
\toprule
\multirow{2}{*}{\textbf{Model Config}} & \multicolumn{2}{c|}{\textbf{Vista}} & \multicolumn{2}{c|}{\textbf{SPAQ}} & \multicolumn{2}{c|}{\textbf{KADID}} & \multicolumn{2}{c}{\textbf{KonIQ}} \\
& \textbf{SRCC} & \textbf{PLCC} & \textbf{SRCC} & \textbf{PLCC} & \textbf{SRCC} & \textbf{PLCC} & \textbf{SRCC} & \textbf{PLCC} \\
\midrule
\textbf{Stage 1 Only (Ours, Pairwise)} & \textbf{0.352} & \textbf{0.404} & 0.760 & 0.780 & 0.700 & 0.725 & 0.650 & 0.680 \\
Stage 1 Only (VisualQuality-R1) & 0.327 & 0.375 & 0.792 & 0.815 & \textbf{0.735} & \textbf{0.758} & 0.684 & 0.712 \\
Stage 1 Only (SFT) & 0.312 & 0.355 & \textbf{0.805} & \textbf{0.822} & 0.720 & 0.741 & \textbf{0.695} & \textbf{0.721} \\
\bottomrule
\end{tabular}%
}
\end{table*}

To further verify this, we sampled 500 images from Vista-Bench containing localized blurring in key regions. VisualQuality-R1 significantly overestimated the quality with an average score of 3.94, whereas Q-Probe yielded an average score of 2.06, which is much closer to the human Ground Truth (GT) average score of 1.82.

\subsection{Training Stability: Uncertainty-Aware Alignment vs. Bradley-Terry}
Due to the inherent subjectivity of IQA, relying solely on single-sample ranking losses (e.g., the Bradley-Terry model) makes Stage-1 RL susceptible to noise overfitting on ambiguous samples. Q-Probe mitigates this via a $K$-rollout uncertainty-aware alignment method, utilizing multi-sample variance to downweight gradients for high-uncertainty samples.

Table \ref{tab:rl_modeling} demonstrates that our approach outperforms the Bradley-Terry model across all datasets. Furthermore, Table \ref{tab:variance_trajectory} tracks the average standard deviation of predicted scores across training steps. While the Bradley-Terry model hits an optimization bottleneck and its variance fluctuates (around 0.15--0.20) as it overfits to subjective noise, Q-Probe demonstrates a continuous and stable reduction in variance, confirming superior training stability.

\begin{table*}[h]
\centering
\caption{Performance Comparison of RL Modeling Strategies}
\label{tab:rl_modeling}
\resizebox{\textwidth}{!}{%
\begin{tabular}{l | cc | cc | c | c}
\toprule
\textbf{Model Config} & \textbf{Vista (SRCC)} & \textbf{Vista (PLCC)} & \textbf{SPAQ (SRCC)} & \textbf{SPAQ (PLCC)} & \textbf{KADID (SRCC)} & \textbf{KonIQ (SRCC)} \\
\midrule
\textbf{Stage 1 Only (Ours)} & \textbf{0.352} & \textbf{0.404} & \textbf{0.760} & \textbf{0.780} & \textbf{0.700} & \textbf{0.650} \\
Stage 1 Only (Bradley-Terry) & 0.337 & 0.388 & 0.745 & 0.762 & 0.682 & 0.635 \\
\bottomrule
\end{tabular}%
}
\end{table*}

\begin{table*}[h]
\centering
\caption{Variance Trajectory During Stage-1 Training. \textcolor{blue}{Blue} values represent our method (Ours), and \textcolor{gray}{gray} values represent the Bradley-Terry (BT) baseline.}
\label{tab:variance_trajectory}
\begin{tabular}{c | cc | cc | cc}
\toprule
\textbf{Training Step} & \multicolumn{2}{c|}{\textbf{Vista}} & \multicolumn{2}{c|}{\textbf{KADID}} & \multicolumn{2}{c}{\textbf{KonIQ}} \\
\midrule
0 & \textcolor{blue}{0.34} & \textcolor{gray}{0.35} & \textcolor{blue}{0.15} & \textcolor{gray}{0.15} & \textcolor{blue}{0.25} & \textcolor{gray}{0.25} \\
50 & \textcolor{blue}{0.18} & \textcolor{gray}{0.26} & \textcolor{blue}{0.08} & \textcolor{gray}{0.12} & \textcolor{blue}{0.10} & \textcolor{gray}{0.22} \\
100 & \textcolor{blue}{0.11} & \textcolor{gray}{0.22} & \textcolor{blue}{0.04} & \textcolor{gray}{0.10} & \textcolor{blue}{0.06} & \textcolor{gray}{0.18} \\
200 & \textcolor{blue}{\textbf{0.08}} & \textcolor{gray}{0.21} & \textcolor{blue}{\textbf{0.01}} & \textcolor{gray}{0.11} & \textcolor{blue}{\textbf{0.01}} & \textcolor{gray}{0.16} \\
\bottomrule
\end{tabular}
\end{table*}

\subsection{Localization Precision and SFT Trade-offs}
\label{tradeoff}
During the Stage-2 SFT phase, we intentionally relax rigid constraints on localization precision to allow the model to learn a mix of degradation and background regions. This prevents the model from developing a spurious causal bias where ``cropping implies degradation.'' As shown in Table \ref{tab:localization_tradeoff}, this strategy slightly trades off localization accuracy (mIoU) for substantial gains in logical robustness (SRCC/PLCC). Stage-3 RL optimization subsequently recovers and maximizes the mIoU via a dedicated location reward, achieving optimal fine-grained detection.

\begin{table}[h]
\centering
\caption{Trade-off Analysis for Localization Precision. Config A/B do not contain Stage-3.}
\label{tab:localization_tradeoff}
\resizebox{\textwidth}{!}{%
\begin{tabular}{l | l | c | c | c}
\toprule
\textbf{Config} & \textbf{Training Strategy} & \textbf{mIoU} & \textbf{SRCC} & \textbf{PLCC} \\
\midrule
Config A & SFT (Strict degradation-only regions) & 0.689 & 0.558 & 0.627 \\
Config B & SFT (Mix of degradation \& background) & 0.643 & 0.762 & 0.808 \\
Config C & \textbf{Full Pipeline (Stage-3 RL Optimization)} & \textbf{0.702} & \textbf{0.801} & \textbf{0.850} \\
\bottomrule
\end{tabular}%
}
\end{table}

\subsection{Reward Parameter Analysis in Decoupled Post-RL}
\label{rewardAbl}
We conducted ablations on the reward parameters in Stage 3: $\alpha$ (accuracy), $\beta$ (localization), and $\gamma$ (format). The temperature parameter $\tau$ in the accuracy reward was empirically set to $1.0$. As demonstrated in Table \ref{tab:reward_ablation}, lowering the weight of either the accuracy or localization reward strictly diminishes performance. The format reward has a marginal influence, while accuracy and localization rewards are the primary drivers of Q-Probe's precision.

\begin{table}[h]
\centering
\caption{Parameter Ablation for Decoupled Rewards}
\label{tab:reward_ablation}
\begin{tabular}{c c c | c c}
\toprule
\boldmath{$\alpha$} \textbf{(Accuracy)} & \boldmath{$\beta$} \textbf{(Localization)} & \boldmath{$\gamma$} \textbf{(Format)} & \textbf{SRCC} & \textbf{PLCC} \\
\midrule
\textbf{1.0} & \textbf{1.0} & \textbf{1.0} & \textbf{0.801} & \textbf{0.850} \\
0.5 & 1.0 & 1.0 & 0.768 & 0.794 \\
1.0 & 0.5 & 1.0 & 0.782 & 0.796 \\
1.0 & 1.0 & 0.5 & 0.799 & 0.851 \\
\bottomrule
\end{tabular}
\end{table}

\subsection{Rigorous Human Annotation and Ground Truth Validation}
To guarantee the absolute reliability of the Vista-Bench evaluation, we emphasize that the final Ground Truth (GT) labels are derived entirely from rigorous manual annotation from scratch.

Given the inherently subjective nature of Image Quality Assessment, we eschewed any reliance on automated systems or large language models for scoring. Instead, we established a dedicated panel of domain experts to conduct systematic, cross-verified evaluations of every single image in the dataset. A sample was strictly discarded if it triggered either:
\begin{enumerate}
    \item \textbf{High Scoring Ambiguity:} The variance among the human evaluators' independently assigned scores was too large, indicating a lack of human consensus.
    \item \textbf{Semantic Disagreement:} The experts failed to reach a consensus on the semantic classification of the cropped regions (e.g., disagreeing on whether a region constituted Primary Foreground or Background).
\end{enumerate}

Through this purely manual, expert-driven annotation process, all data splits are fully vetted and approved by human evaluators, successfully capturing the plurality of human opinions while ensuring the highest standard of data fidelity.

\newpage
\section*{NeurIPS Paper Checklist}

\begin{enumerate}

\item {\bf Claims}
    \item[] Question: Do the main claims made in the abstract and introduction accurately reflect the paper's contributions and scope?
    \item[] Answer: \answerYes{}
    \item[] Justification: Our abstract and introduction clearly outline our contributions, including the Vista-Bench benchmark, the Q-Probe framework, and the three-stage training paradigm, all of which are extensively supported by our methodology and experiments sections.
    \item[] Guidelines:
    \begin{itemize}
        \item The answer \answerNA{} means that the abstract and introduction do not include the claims made in the paper.
        \item The abstract and/or introduction should clearly state the claims made, including the contributions made in the paper and important assumptions and limitations. A \answerNo{} or \answerNA{} answer to this question will not be perceived well by the reviewers. 
        \item The claims made should match theoretical and experimental results, and reflect how much the results can be expected to generalize to other settings. 
        \item It is fine to include aspirational goals as motivation as long as it is clear that these goals are not attained by the paper. 
    \end{itemize}

\item {\bf Limitations}
    \item[] Question: Does the paper discuss the limitations of the work performed by the authors?
    \item[] Answer: \answerYes{}
    \item[] Justification: We include a dedicated "Limitations and Bad Cases Study" section (Appendix E) that discusses and visualizes semantic entanglement, spatial context insensitivity, and salience bias.
    \item[] Guidelines:
    \begin{itemize}
        \item The answer \answerNA{} means that the paper has no limitation while the answer \answerNo{} means that the paper has limitations, but those are not discussed in the paper. 
        \item The authors are encouraged to create a separate ``Limitations'' section in their paper.
        \item The paper should point out any strong assumptions and how robust the results are to violations of these assumptions (e.g., independence assumptions, noiseless settings, model well-specification, asymptotic approximations only holding locally). The authors should reflect on how these assumptions might be violated in practice and what the implications would be.
        \item The authors should reflect on the scope of the claims made, e.g., if the approach was only tested on a few datasets or with a few runs. In general, empirical results often depend on implicit assumptions, which should be articulated.
        \item The authors should reflect on the factors that influence the performance of the approach. For example, a facial recognition algorithm may perform poorly when image resolution is low or images are taken in low lighting. Or a speech-to-text system might not be used reliably to provide closed captions for online lectures because it fails to handle technical jargon.
        \item The authors should discuss the computational efficiency of the proposed algorithms and how they scale with dataset size.
        \item If applicable, the authors should discuss possible limitations of their approach to address problems of privacy and fairness.
        \item While the authors might fear that complete honesty about limitations might be used by reviewers as grounds for rejection, a worse outcome might be that reviewers discover limitations that aren't acknowledged in the paper. The authors should use their best judgment and recognize that individual actions in favor of transparency play an important role in developing norms that preserve the integrity of the community. Reviewers will be specifically instructed to not penalize honesty concerning limitations.
    \end{itemize}

\item {\bf Theory assumptions and proofs}
    \item[] Question: For each theoretical result, does the paper provide the full set of assumptions and a complete (and correct) proof?
    \item[] Answer: \answerNA{}
    \item[] Justification: Our work is primarily empirical and methodological, focusing on an agentic IQA framework and a new dataset benchmark, and we do not present formal theoretical proofs.
    \item[] Guidelines:
    \begin{itemize}
        \item The answer \answerNA{} means that the paper does not include theoretical results. 
        \item All the theorems, formulas, and proofs in the paper should be numbered and cross-referenced.
        \item All assumptions should be clearly stated or referenced in the statement of any theorems.
        \item The proofs can either appear in the main paper or the supplemental material, but if they appear in the supplemental material, the authors are encouraged to provide a short proof sketch to provide intuition. 
        \item Inversely, any informal proof provided in the core of the paper should be complemented by formal proofs provided in appendix or supplemental material.
        \item Theorems and Lemmas that the proof relies upon should be properly referenced. 
    \end{itemize}

    \item {\bf Experimental result reproducibility}
    \item[] Question: Does the paper fully disclose all the information needed to reproduce the main experimental results of the paper to the extent that it affects the main claims and/or conclusions of the paper (regardless of whether the code and data are provided or not)?
    \item[] Answer: \answerYes{}
    \item[] Justification: We provide comprehensive details on our experimental setup, including base models (Qwen-2.5-VL-7B), hyperparameter settings (e.g., GRPO generation number, learning rate), evaluation metrics, and system prompts used for trajectory generation and inference.
    \item[] Guidelines:
    \begin{itemize}
        \item The answer \answerNA{} means that the paper does not include experiments.
        \item If the paper includes experiments, a \answerNo{} answer to this question will not be perceived well by the reviewers: Making the paper reproducible is important, regardless of whether the code and data are provided or not.
        \item If the contribution is a dataset and\slash or model, the authors should describe the steps taken to make their results reproducible or verifiable. 
        \item Depending on the contribution, reproducibility can be accomplished in various ways. For example, if the contribution is a novel architecture, describing the architecture fully might suffice, or if the contribution is a specific model and empirical evaluation, it may be necessary to either make it possible for others to replicate the model with the same dataset, or provide access to the model. In general. releasing code and data is often one good way to accomplish this, but reproducibility can also be provided via detailed instructions for how to replicate the results, access to a hosted model (e.g., in the case of a large language model), releasing of a model checkpoint, or other means that are appropriate to the research performed.
        \item While NeurIPS does not require releasing code, the conference does require all submissions to provide some reasonable avenue for reproducibility, which may depend on the nature of the contribution. For example
        \begin{enumerate}
            \item If the contribution is primarily a new algorithm, the paper should make it clear how to reproduce that algorithm.
            \item If the contribution is primarily a new model architecture, the paper should describe the architecture clearly and fully.
            \item If the contribution is a new model (e.g., a large language model), then there should either be a way to access this model for reproducing the results or a way to reproduce the model (e.g., with an open-source dataset or instructions for how to construct the dataset).
            \item We recognize that reproducibility may be tricky in some cases, in which case authors are welcome to describe the particular way they provide for reproducibility. In the case of closed-source models, it may be that access to the model is limited in some way (e.g., to registered users), but it should be possible for other researchers to have some path to reproducing or verifying the results.
        \end{enumerate}
    \end{itemize}

\item {\bf Open access to data and code}
    \item[] Question: Does the paper provide open access to the data and code, with sufficient instructions to faithfully reproduce the main experimental results, as described in supplemental material?
    \item[] Answer: \answerYes{}
    \item[] Justification: We have included the code and dataset in the supplementary materials. Furthermore, we provide detailed descriptions in both the main text and the appendix on how the code is implemented and how the dataset is constructed (including the ground-truth scores and Chain-of-Thought (CoT) reasoning).
    \item[] Guidelines:
    \begin{itemize}
        \item The answer \answerNA{} means that paper does not include experiments requiring code.
        \item Please see the NeurIPS code and data submission guidelines (\url{https://neurips.cc/public/guides/CodeSubmissionPolicy}) for more details.
        \item While we encourage the release of code and data, we understand that this might not be possible, so \answerNo{} is an acceptable answer. Papers cannot be rejected simply for not including code, unless this is central to the contribution (e.g., for a new open-source benchmark).
        \item The instructions should contain the exact command and environment needed to run to reproduce the results. See the NeurIPS code and data submission guidelines (\url{https://neurips.cc/public/guides/CodeSubmissionPolicy}) for more details.
        \item The authors should provide instructions on data access and preparation, including how to access the raw data, preprocessed data, intermediate data, and generated data, etc.
        \item The authors should provide scripts to reproduce all experimental results for the new proposed method and baselines. If only a subset of experiments are reproducible, they should state which ones are omitted from the script and why.
        \item At submission time, to preserve anonymity, the authors should release anonymized versions (if applicable).
        \item Providing as much information as possible in supplemental material (appended to the paper) is recommended, but including URLs to data and code is permitted.
    \end{itemize}

\item {\bf Experimental setting/details}
    \item[] Question: Does the paper specify all the training and test details (e.g., data splits, hyperparameters, how they were chosen, type of optimizer) necessary to understand the results?
    \item[] Answer: \answerYes{}
    \item[] Justification: In Section 4.1, we specify key training parameters, including the AdamW optimizer, learning rate, GRPO generation numbers, clip threshold, and batch/crop sizing.
    \item[] Guidelines:
    \begin{itemize}
        \item The answer \answerNA{} means that the paper does not include experiments.
        \item The experimental setting should be presented in the core of the paper to a level of detail that is necessary to appreciate the results and make sense of them.
        \item The full details can be provided either with the code, in appendix, or as supplemental material.
    \end{itemize}

\item {\bf Experiment statistical significance}
    \item[] Question: Does the paper report error bars suitably and correctly defined or other appropriate information about the statistical significance of the experiments?
    \item[] Answer: \answerYes{}
    \item[] Justification: We track the average standard deviation of predicted scores across multiple inference runs over various training steps (Figure 4, Table 9) to empirically demonstrate our training stability and variance reduction.
    \item[] Guidelines:
    \begin{itemize}
        \item The answer \answerNA{} means that the paper does not include experiments.
        \item The authors should answer \answerYes{} if the results are accompanied by error bars, confidence intervals, or statistical significance tests, at least for the experiments that support the main claims of the paper.
        \item The factors of variability that the error bars are capturing should be clearly stated (for example, train/test split, initialization, random drawing of some parameter, or overall run with given experimental conditions).
        \item The method for calculating the error bars should be explained (closed form formula, call to a library function, bootstrap, etc.)
        \item The assumptions made should be given (e.g., Normally distributed errors).
        \item It should be clear whether the error bar is the standard deviation or the standard error of the mean.
        \item It is OK to report 1-sigma error bars, but one should state it. The authors should preferably report a 2-sigma error bar than state that they have a 96\% CI, if the hypothesis of Normality of errors is not verified.
        \item For asymmetric distributions, the authors should be careful not to show in tables or figures symmetric error bars that would yield results that are out of range (e.g., negative error rates).
        \item If error bars are reported in tables or plots, the authors should explain in the text how they were calculated and reference the corresponding figures or tables in the text.
    \end{itemize}

\item {\bf Experiments compute resources}
    \item[] Question: For each experiment, does the paper provide sufficient information on the computer resources (type of compute workers, memory, time of execution) needed to reproduce the experiments?
    \item[] Answer: \answerYes{}
    \item[] Justification: We explicitly state that our experiments were conducted on 8 NVIDIA A100 (80GB) GPUs and that we deployed the inference pipeline across NVIDIA RTX 4090 GPUs.
    \item[] Guidelines:
    \begin{itemize}
        \item The answer \answerNA{} means that the paper does not include experiments.
        \item The paper should indicate the type of compute workers CPU or GPU, internal cluster, or cloud provider, including relevant memory and storage.
        \item The paper should provide the amount of compute required for each of the individual experimental runs as well as estimate the total compute. 
        \item The paper should disclose whether the full research project required more compute than the experiments reported in the paper (e.g., preliminary or failed experiments that didn't make it into the paper). 
    \end{itemize}
    
\item {\bf Code of ethics}
    \item[] Question: Does the research conducted in the paper conform, in every respect, with the NeurIPS Code of Ethics \url{https://neurips.cc/public/EthicsGuidelines}?
    \item[] Answer: \answerYes{}
    \item[] Justification: Our research focuses on advancing image quality assessment methodology safely and does not pose any obvious ethical violations, malicious use cases, or involve harmful data processing.
    \item[] Guidelines:
    \begin{itemize}
        \item The answer \answerNA{} means that the authors have not reviewed the NeurIPS Code of Ethics.
        \item If the authors answer \answerNo, they should explain the special circumstances that require a deviation from the Code of Ethics.
        \item The authors should make sure to preserve anonymity (e.g., if there is a special consideration due to laws or regulations in their jurisdiction).
    \end{itemize}

\item {\bf Broader impacts}
    \item[] Question: Does the paper discuss both potential positive societal impacts and negative societal impacts of the work performed?
    \item[] Answer: \answerNA{}{}
    \item[] Justification: We focus intensely on the technical methodology of high-resolution IQA and this won't have potential positive societal impacts or negative societal impacts.
    \item[] Guidelines:
    \begin{itemize}
        \item The answer \answerNA{} means that there is no societal impact of the work performed.
        \item If the authors answer \answerNA{} or \answerNo, they should explain why their work has no societal impact or why the paper does not address societal impact.
        \item Examples of negative societal impacts include potential malicious or unintended uses (e.g., disinformation, generating fake profiles, surveillance), fairness considerations (e.g., deployment of technologies that could make decisions that unfairly impact specific groups), privacy considerations, and security considerations.
        \item The conference expects that many papers will be foundational research and not tied to particular applications, let alone deployments. However, if there is a direct path to any negative applications, the authors should point it out. For example, it is legitimate to point out that an improvement in the quality of generative models could be used to generate Deepfakes for disinformation. On the other hand, it is not needed to point out that a generic algorithm for optimizing neural networks could enable people to train models that generate Deepfakes faster.
        \item The authors should consider possible harms that could arise when the technology is being used as intended and functioning correctly, harms that could arise when the technology is being used as intended but gives incorrect results, and harms following from (intentional or unintentional) misuse of the technology.
        \item If there are negative societal impacts, the authors could also discuss possible mitigation strategies (e.g., gated release of models, providing defenses in addition to attacks, mechanisms for monitoring misuse, mechanisms to monitor how a system learns from feedback over time, improving the efficiency and accessibility of ML).
    \end{itemize}
    
\item {\bf Safeguards}
    \item[] Question: Does the paper describe safeguards that have been put in place for responsible release of data or models that have a high risk for misuse (e.g., pre-trained language models, image generators, or scraped datasets)?
    \item[] Answer: \answerNA{}
    \item[] Justification: Our proposed Q-Probe framework and Vista-Bench evaluate technical image fidelity and distortions, which present a very low risk for misuse compared to generative models or sensitive demographic datasets.
    \item[] Guidelines:
    \begin{itemize}
        \item The answer \answerNA{} means that the paper poses no such risks.
        \item Released models that have a high risk for misuse or dual-use should be released with necessary safeguards to allow for controlled use of the model, for example by requiring that users adhere to usage guidelines or restrictions to access the model or implementing safety filters. 
        \item Datasets that have been scraped from the Internet could pose safety risks. The authors should describe how they avoided releasing unsafe images.
        \item We recognize that providing effective safeguards is challenging, and many papers do not require this, but we encourage authors to take this into account and make a best faith effort.
    \end{itemize}

\item {\bf Licenses for existing assets}
    \item[] Question: Are the creators or original owners of assets (e.g., code, data, models), used in the paper, properly credited and are the license and terms of use explicitly mentioned and properly respected?
    \item[] Answer: \answerYes{}
    \item[] Justification: We appropriately credit and cite existing datasets (HR-Bench 4K, UHD-IQA, KADID-10k) and foundation models (Qwen-2.5-VL-7B), adhering to academic standards, though we do not print specific license texts inline.
    \item[] Guidelines:
    \begin{itemize}
        \item The answer \answerNA{} means that the paper does not use existing assets.
        \item The authors should cite the original paper that produced the code package or dataset.
        \item The authors should state which version of the asset is used and, if possible, include a URL.
        \item The name of the license (e.g., CC-BY 4.0) should be included for each asset.
        \item For scraped data from a particular source (e.g., website), the copyright and terms of service of that source should be provided.
        \item If assets are released, the license, copyright information, and terms of use in the package should be provided. For popular datasets, \url{paperswithcode.com/datasets} has curated licenses for some datasets. Their licensing guide can help determine the license of a dataset.
        \item For existing datasets that are re-packaged, both the original license and the license of the derived asset (if it has changed) should be provided.
        \item If this information is not available online, the authors are encouraged to reach out to the asset's creators.
    \end{itemize}

\item {\bf New assets}
    \item[] Question: Are new assets introduced in the paper well documented and is the documentation provided alongside the assets?
    \item[] Answer: \answerYes{}
    \item[] Justification: We extensively detail our Vista-Bench construction, including our wavelet-based degradation methodology, annotation procedures, and SFT format guidelines (Probe-CoT-3K), in both the main text and appendices.
    \item[] Guidelines:
    \begin{itemize}
        \item The answer \answerNA{} means that the paper does not release new assets.
        \item Researchers should communicate the details of the dataset\slash code\slash model as part of their submissions via structured templates. This includes details about training, license, limitations, etc. 
        \item The paper should discuss whether and how consent was obtained from people whose asset is used.
        \item At submission time, remember to anonymize your assets (if applicable). You can either create an anonymized URL or include an anonymized zip file.
    \end{itemize}

\item {\bf Crowdsourcing and research with human subjects}
    \item[] Question: For crowdsourcing experiments and research with human subjects, does the paper include the full text of instructions given to participants and screenshots, if applicable, as well as details about compensation (if any)? 
    \item[] Answer: \answerYes{}
    \item[] Justification: We detail the human annotation process and instructions in Appendix A.2. Furthermore, all six specialists (domain experts and PhD researchers) were fairly compensated for their annotation workload through our laboratory's research funding at standard institutional rates, adhering to ethical labor practices.
    \item[] Guidelines:
    \begin{itemize}
        \item The answer \answerNA{} means that the paper does not involve crowdsourcing nor research with human subjects.
        \item Including this information in the supplemental material is fine, but if the main contribution of the paper involves human subjects, then as much detail as possible should be included in the main paper. 
        \item According to the NeurIPS Code of Ethics, workers involved in data collection, curation, or other labor should be paid at least the minimum wage in the country of the data collector. 
    \end{itemize}

\item {\bf Institutional review board (IRB) approvals or equivalent for research with human subjects}
    \item[] Question: Does the paper describe potential risks incurred by study participants, whether such risks were disclosed to the subjects, and whether Institutional Review Board (IRB) approvals (or an equivalent approval/review based on the requirements of your country or institution) were obtained?
    \item[] Answer: \answerNA{}
    \item[] Justification: Our "human subjects" involvement was restricted to expert evaluators grading the technical quality of images, which is standard data annotation lacking the risk or intervention that typically requires IRB oversight.
    \item[] Guidelines:
    \begin{itemize}
        \item The answer \answerNA{} means that the paper does not involve crowdsourcing nor research with human subjects.
        \item Depending on the country in which research is conducted, IRB approval (or equivalent) may be required for any human subjects research. If you obtained IRB approval, you should clearly state this in the paper. 
        \item We recognize that the procedures for this may vary significantly between institutions and locations, and we expect authors to adhere to the NeurIPS Code of Ethics and the guidelines for their institution. 
        \item For initial submissions, do not include any information that would break anonymity (if applicable), such as the institution conducting the review.
    \end{itemize}

\item {\bf Declaration of LLM usage}
    \item[] Question: Does the paper describe the usage of LLMs if it is an important, original, or non-standard component of the core methods in this research? Note that if the LLM is used only for writing, editing, or formatting purposes and does \emph{not} impact the core methodology, scientific rigor, or originality of the research, declaration is not required.
    \item[] Answer: \answerYes{}
    \item[] Justification: We explicitly state our use of an MLLM (Qwen-2.5-VL-7B) as the foundational architecture for our Q-Probe framework and for autonomous trajectory synthesis within our methodology section.
    \item[] Guidelines:
    \begin{itemize}
        \item The answer \answerNA{} means that the core method development in this research does not involve LLMs as any important, original, or non-standard components.
        \item Please refer to our LLM policy in the NeurIPS handbook for what should or should not be described.
    \end{itemize}

\end{enumerate}
\end{document}